\pdfoutput=1
\documentclass[12pt,a4paper]{article}

\usepackage{ifthen} 
\newboolean{pdflatex}
\setboolean{pdflatex}{true} 

\newboolean{articletitles}
\setboolean{articletitles}{true} 

\newboolean{uprightparticles}
\setboolean{uprightparticles}{false} 

\def\paperauthors{LHCb collaboration} 
\def\paperasciititle{Template for writing LHCb papers} 
\def\papertitle{Differential decay rate of $B^+ \to \Jpsi K^+$ with the LHCb Upgrade I experiment} 
\def\paperkeywords{{High Energy Physics}, {LHCb}} 
\def\papercopyright{\the\year\ CERN for the benefit of the LHCb collaboration} 
\def\paperlicence{CC BY 4.0 licence}
\def\paperlicenceurl{https://creativecommons.org/licenses/by/4.0/}

\newif\ifEnableSectionTOCLinks
\EnableSectionTOCLinksfalse 

\usepackage[top=1in, bottom=1.25in, left=1in, right=1in]{geometry}

\usepackage{microtype}
\usepackage{lineno}  
\usepackage{xspace} 
\usepackage{caption} 

\usepackage{graphicx}  
\usepackage{color}
\usepackage{colortbl}
\graphicspath{{./figs/}} 

\usepackage{amsmath} 
\usepackage{amssymb}
\usepackage{amsfonts}
\usepackage{upgreek} 

\newcommand*\patchAmsMathEnvironmentForLineno[1]{%
\expandafter\let\csname old#1\expandafter\endcsname\csname #1\endcsname
\expandafter\let\csname oldend#1\expandafter\endcsname\csname
end#1\endcsname
 \renewenvironment{#1}%
   {\linenomath\csname old#1\endcsname}%
   {\csname oldend#1\endcsname\endlinenomath}%
}
\newcommand*\patchBothAmsMathEnvironmentsForLineno[1]{%
  \patchAmsMathEnvironmentForLineno{#1}%
  \patchAmsMathEnvironmentForLineno{#1*}%
}
\AtBeginDocument{%
\patchBothAmsMathEnvironmentsForLineno{equation}%
\patchBothAmsMathEnvironmentsForLineno{align}%
\patchBothAmsMathEnvironmentsForLineno{flalign}%
\patchBothAmsMathEnvironmentsForLineno{alignat}%
\patchBothAmsMathEnvironmentsForLineno{gather}%
\patchBothAmsMathEnvironmentsForLineno{multline}%
\patchBothAmsMathEnvironmentsForLineno{eqnarray}%
}

\usepackage[pdftex,
            pdfauthor={\paperauthors},
            pdftitle={\paperasciititle},
            pdfkeywords={\paperkeywords}]{hyperref}
\usepackage{hyperxmp}
\hypersetup{
    pdfcopyright={Copyright (C) \papercopyright},
    pdflicenseurl={\paperlicenceurl}
}

\usepackage[colorinlistoftodos,textsize=scriptsize]{todonotes}

\usepackage[bottom,flushmargin,hang,multiple]{footmisc}

\usepackage[all]{hypcap} 

\usepackage{xspace}
\usepackage{upgreek}

\def\lhcb   {\mbox{LHCb}\xspace}

\def\MagUp {\mbox{\em Mag\kern -0.05em Up}\xspace}
\def\MagDown {\mbox{\em MagDown}\xspace}

\def\hltone {HLT1\xspace}
\def\hlttwo {HLT2\xspace}

\ifthenelse{\boolean{uprightparticles}}%
{

 \def\Pmu         {\ensuremath{\upmu}\xspace}

 \def\Ppi         {\ensuremath{\uppi}\xspace}

 \def\Ppsi        {\ensuremath{\uppsi}\xspace}

 \def\PDelta      {\ensuremath{\Delta}\xspace}
 \def\PXi         {\ensuremath{\Xi}\xspace}
 \def\PLambda     {\ensuremath{\Lambda}\xspace}
 \def\PSigma      {\ensuremath{\Sigma}\xspace}
 \def\POmega      {\ensuremath{\Omega}\xspace}
 \def\PUpsilon    {\ensuremath{\Upsilon}\xspace}
 \let\oldPi\Pi
 \def\PPi         {\ensuremath{\oldPi}\xspace}

 \def\PB      {\ensuremath{\mathrm{B}}\xspace}
 \def\PD      {\ensuremath{\mathrm{D}}\xspace}
 \def\PJ      {\ensuremath{\mathrm{J}}\xspace}
 \def\PK      {\ensuremath{\mathrm{K}}\xspace}
 \def\Pb      {\ensuremath{\mathrm{b}}\xspace}
 \def\Pc      {\ensuremath{\mathrm{c}}\xspace}

 \def\Ps      {\ensuremath{\mathrm{s}}\xspace}

 \def\thebaroffset{0.0em}
}
{

 \def\Pmu         {\ensuremath{\mu}\xspace}

 \def\Ppi         {\ensuremath{\pi}\xspace}

 \def\Ppsi        {\ensuremath{\psi}\xspace}
 
 \mathchardef\PDelta="7101
 \mathchardef\PXi="7104
 \mathchardef\PLambda="7103
 \mathchardef\PSigma="7106
 \mathchardef\POmega="710A
 \mathchardef\PUpsilon="7107
 \mathchardef\PPi="7105
 \def\PB      {\ensuremath{B}\xspace}
 \def\PD      {\ensuremath{D}\xspace}
 \def\PJ      {\ensuremath{J}\xspace}
 \def\PK      {\ensuremath{K}\xspace}
 \def\Pb      {\ensuremath{b}\xspace}
 \def\Pc      {\ensuremath{c}\xspace}

 \def\Ps      {\ensuremath{s}\xspace}

 \def\thebaroffset{0.18em}
}
\newcommand{\offsetoverline}[2][\thebaroffset]{\kern #1\overline{\kern -#1 #2}}%

\makeatletter
\ifcase \@ptsize \relax
  \newcommand{\miniscule}{\@setfontsize\miniscule{4}{5}}
\or
  \newcommand{\miniscule}{\@setfontsize\miniscule{5}{6}}
\or
  \newcommand{\miniscule}{\@setfontsize\miniscule{5}{6}}
\fi
\makeatother

\DeclareRobustCommand{\optbar}[1]{\shortstack{{\miniscule (\rule[.5ex]{1.25em}{.18mm})}
  \\ [-.7ex] $#1$}}

\def\mup        {{\ensuremath{\Pmu^+}}\xspace}
\def\mun        {{\ensuremath{\Pmu^-}}\xspace} 

\def\squark    {{\ensuremath{\Ps}}\xspace}

\def\cquark    {{\ensuremath{\Pc}}\xspace}
\def\cquarkbar {{\ensuremath{\overline \cquark}}\xspace}

\def\bquark    {{\ensuremath{\Pb}}\xspace}

\def\pion   {{\ensuremath{\Ppi}}\xspace}

\def\pim    {{\ensuremath{\pion^-}}\xspace}

\def\kaon    {{\ensuremath{\PK}}\xspace}

\def\KorKbar {\kern \thebaroffset\optbar{\kern -\thebaroffset \PK}{}\xspace}

\def\Kp      {{\ensuremath{\kaon^+}}\xspace}

\def\D       {{\ensuremath{\PD}}\xspace}

\def\DorDbar {\kern \thebaroffset\optbar{\kern -\thebaroffset \PD}\xspace}

\def\Dp      {{\ensuremath{\D^+}}\xspace}
\def\Dm      {{\ensuremath{\D^-}}\xspace}

\def\DpDm    {\ensuremath{\Dp {\kern -0.16em \Dm}}\xspace}

\def\B       {{\ensuremath{\PB}}\xspace}

\def\BorBbar {\kern \thebaroffset\optbar{\kern -\thebaroffset \PB}\xspace}

\def\Bd      {{\ensuremath{\B^0}}\xspace}

\def\BdorBdbar {\kern \thebaroffset\optbar{\kern -\thebaroffset \Bd}\xspace}
\def\Bu      {{\ensuremath{\B^+}}\xspace}

\def\Bs      {{\ensuremath{\B^0_\squark}}\xspace}

\def\BsorBsbar {\kern \thebaroffset\optbar{\kern -\thebaroffset \Bs}\xspace}

\def\jpsi     {{\ensuremath{{\PJ\mskip -3mu/\mskip -2mu\Ppsi}}}\xspace}

\def\Y#1S{\ensuremath{\PUpsilon{(#1S)}}\xspace}

\def\LorLbar     {\kern \thebaroffset\optbar{\kern -\thebaroffset \PLambda}\xspace}

\newcommand{\decay}[2]{\mbox{\ensuremath{#1\!\to #2}}\xspace}

\def\to                 {\ensuremath{\rightarrow}\xspace}

\def\qsq       {{\ensuremath{q^2}}\xspace}

\def\AT#1     {\ensuremath{A_{\mathrm{T}}^{#1}}\xspace}           

\def\ctl       {\ensuremath{\cos{\theta_\ell}}\xspace}

\def\C#1      {\ensuremath{\mathcal{C}_{#1}}\xspace}                       
\def\Cp#1     {\ensuremath{\mathcal{C}_{#1}^{'}}\xspace}                    
\def\Ceff#1   {\ensuremath{\mathcal{C}_{#1}^{\mathrm{(eff)}}}\xspace}        
\def\Cpeff#1  {\ensuremath{\mathcal{C}_{#1}^{'\mathrm{(eff)}}}\xspace}       
\def\Ope#1    {\ensuremath{\mathcal{O}_{#1}}\xspace}                       
\def\Opep#1   {\ensuremath{\mathcal{O}_{#1}^{'}}\xspace}                    

\newcommand{\aunit}[1]{\ensuremath{\text{\,#1}}}

\newcommand{\tev}{\aunit{Te\kern -0.1em V}\xspace}
\newcommand{\gev}{\aunit{Ge\kern -0.1em V}\xspace}
\newcommand{\mev}{\aunit{Me\kern -0.1em V}\xspace}
\newcommand{\kev}{\aunit{ke\kern -0.1em V}\xspace}
\newcommand{\ev}{\aunit{e\kern -0.1em V}\xspace}

\newcommand{\mevc}{\ensuremath{\aunit{Me\kern -0.1em V\!/}c}\xspace}
\newcommand{\gevc}{\ensuremath{\aunit{Ge\kern -0.1em V\!/}c}\xspace}
\newcommand{\mevcc}{\ensuremath{\aunit{Me\kern -0.1em V\!/}c^2}\xspace}
\newcommand{\gevcc}{\ensuremath{\aunit{Ge\kern -0.1em V\!/}c^2}\xspace}
\newcommand{\gevgevcccc}{\ensuremath{\gev^2\!/c^4}\xspace} 

\def\fb   {\ensuremath{\aunit{fb}}\xspace}
\def\invfb   {\ensuremath{\fb^{-1}}\xspace}

\def\mhz  {\ensuremath{\aunit{MHz}}\xspace}

\newcommand{\stat}{\aunit{(stat)}\xspace}
\newcommand{\syst}{\aunit{(syst)}\xspace}

\def\gsim{{~\raise.15em\hbox{$>$}\kern-.85em
          \lower.35em\hbox{$\sim$}~}\xspace}
\def\lsim{{~\raise.15em\hbox{$<$}\kern-.85em
          \lower.35em\hbox{$\sim$}~}\xspace}

\def\sPlot{\mbox{\em sPlot}\xspace}

\def\pt         {\ensuremath{p_{\mathrm{T}}}\xspace}

\def\evtgen     {\mbox{\textsc{EvtGen}}\xspace}

\def\geant      {\mbox{\textsc{Geant4}}\xspace}

\def\photos     {\mbox{\textsc{Photos}}\xspace}

\def\pythia     {\mbox{\textsc{Pythia}}\xspace}

\def\tell1  {TELL1\xspace}
\def\ukl1   {UKL1\xspace}

\newcommand{\eg}{\mbox{\itshape e.g.}\xspace}
\newcommand{\ie}{\mbox{\itshape i.e.}\xspace}

\newcommand{\lhcborcid}[1]{\href{https://orcid.org/#1}{\hspace*{0.1em}\raisebox{-0.45ex}{\includegraphics[width=1em]{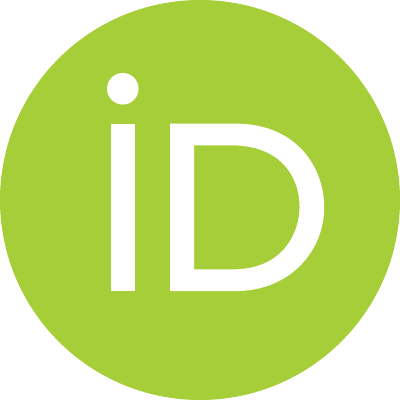}}}}

\hypersetup{
  colorlinks   = true, 
  urlcolor     = blue, 
  linkcolor    = blue, 
  citecolor    = red   
}

\ifEnableSectionTOCLinks
    \usepackage[explicit]{titlesec} 
    
    \let\oldcontentsline\contentsline
    \renewcommand

    \titleformat{\section}{\normalfont\Large\bf}{\hyperlink{tocsection.\thesection}{{\thesection} \parbox[t]{\dimexpr\textwidth-1pc}{#1}}}{1pc}{}

    \titleformat{\subsection}{\normalfont\bf}{\hyperlink{tocsubsection.\thesubsection}{{\thesubsection} \parbox[t]{\dimexpr\textwidth-1pc}{#1}}}{1pc}{}

    \titleformat{name=\section,numberless}[display]{}{}{0pt}{\normalfont\Huge\bfseries #1}
\fi

\usepackage{cite} 
\usepackage{mciteplus}

\usepackage{longtable} 

\def\wkin{\ensuremath{w_{\rm GBR}}\xspace}
\def\Jpsi{\ensuremath{J/\psi}\xspace}
\def\Jpsimm{\decay{\Jpsi}{\mu^+\mu^-}}
\def\afb {\ensuremath{A_{\rm FB}}\xspace}
\def\fh {\ensuremath{F_{\rm H}}\xspace}
\def\btosmumu{\decay{b}{s\mu^+\mu^-}}
\def\btodmumu{\decay{b}{d\mu^+\mu^-}}
\def\KJpsi{\decay{B^+}{\Jpsi K^+}}
\def\PiJpsi{\decay{B^+}{\Jpsi \pi^+}}

\def\mKmm{\ensuremath{m(K^+\mu^+\mu^-)}\xspace}

\def\btosccbar{\decay{\bquark}{\squark \cquarkbar c}}
\def\Kmm{\decay{B^+}{K^+ \mu^+ \mu^-}}
\def\Kstmm{\decay{B^0}{K^{\ast 0} \mu^+ \mu^-}}

\def\Phimm{\decay{B_s^0}{\phi \mu^+ \mu^-}}
\def\Dzpim{\decay{D^{*+}}{D^0(\to K^-\pi^+)\pi^-}}
\def\exoone{\decay{B^0}{T_{c\bar c 1}(4430)^{-}K^+}}

\def\zfit     {\mbox{\textsc{zfit}}\xspace}

\begin{document}

\renewcommand{\thefootnote}{\fnsymbol{footnote}}
\setcounter{footnote}{1}


\begin{titlepage}
\pagenumbering{roman}

\vspace*{-1.5cm}
\centerline{\large EUROPEAN ORGANIZATION FOR NUCLEAR RESEARCH (CERN)}
\vspace*{1.5cm}
\noindent
\begin{tabular*}{\linewidth}{lc@{\extracolsep{\fill}}r@{\extracolsep{0pt}}}
\ifthenelse{\boolean{pdflatex}}
{\vspace*{-1.5cm}\mbox{\!\!\!\includegraphics[width=.14\textwidth]{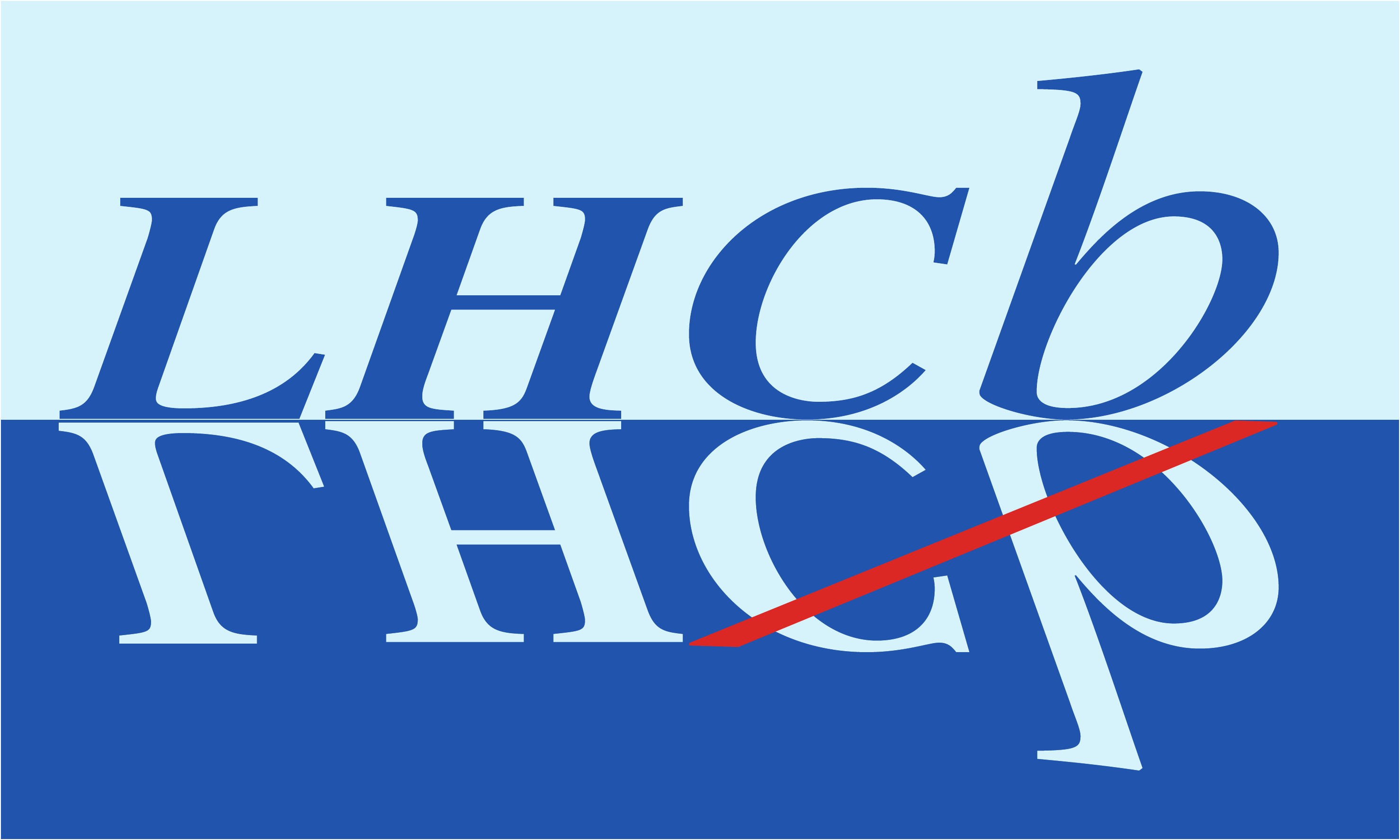}} & &}%
{\vspace*{-1.2cm}\mbox{\!\!\!\includegraphics[width=.12\textwidth]{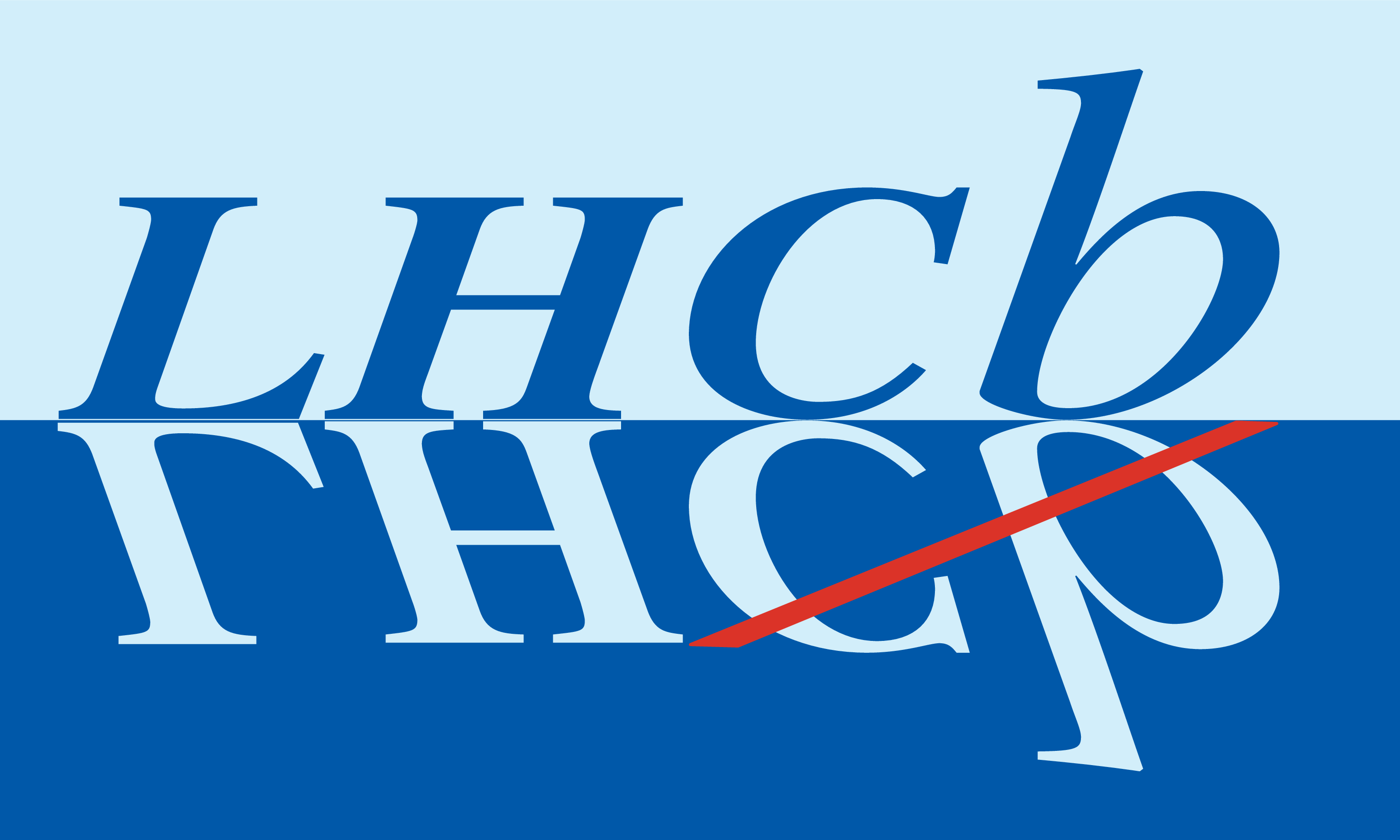}} & &}%
\\
 & & CERN-EP-2025-245 \\  
 & & LHCb-PAPER-2025-040 \\  
 & & June 11, 2025 \\ 
 & & \\
\end{tabular*}

\vspace*{3.0cm}

{\normalfont\bfseries\boldmath\huge
\begin{center}
  \papertitle 
\end{center}
}

\vspace*{2.0cm}

\begin{center}
\paperauthors\footnote{Authors are listed at the end of this paper.}
\end{center}

\vspace{\fill}

\begin{abstract}
\noindent
The normalised decay rate of $B^+ \to J/\psi(\to \mu^+\mu^-) K^+$ is measured as a function of the lepton helicity angle using a data sample corresponding to an integrated luminosity of $1.1 \invfb $ collected
during October 2024 with the upgraded (Upgrade~I) LHCb detector. 
This angular distribution can be parameterised by two coefficients, the forward-backward asymmetry, $A_{\rm FB}$, and the flatness parameter, $F_{\rm H}$, whose values are constrained by conservation of angular momentum.
These coefficients are measured both integrated and differentially across various kinematic and detector-response variables, and the results are found to be in good agreement with expectations. 
These measurements show that the detector response of the LHCb Upgrade I experiment is understood to the precision required to reliably extract the angular coefficients associated with rare $b \to s \mu^+\mu^-$ and $b \to d \mu^+\mu^-$ transitions, which are particularly sensitive to physics beyond the Standard Model.
  
\end{abstract}

\vspace*{2.0cm}

\begin{center}
  Published in Journal of High Energy Physics 04 (2026) 170

\end{center}

\vspace{\fill}

{\footnotesize 
\centerline{\copyright~\papercopyright. \href{\paperlicenceurl}{\paperlicence}.}}
\vspace*{2mm}

\end{titlepage}


\newpage
\setcounter{page}{2}
\mbox{~}
%
%
%
%


\renewcommand{\thefootnote}{\arabic{footnote}}
\setcounter{footnote}{0}


\cleardoublepage


\pagestyle{plain} 
\setcounter{page}{1}
\pagenumbering{arabic}


\section{Introduction}
\label{sec:Introduction}

The \lhcb experiment underwent a major upgrade~\cite{LHCb-DP-2022-002} for the start of LHC Run~3 (2022--2026)  data taking. Most of the detectors have been entirely replaced to enable operation at an instantaneous luminosity five times greater than in previous data-taking periods. Furthermore, all detectors are read out directly into a software trigger system at the LHC crossing rate of 40\mhz, enabling real-time event reconstruction and selection. This upgraded detector is referred to as \lhcb Upgrade~I.
This paper presents a measurement of the normalised differential decay rate of \KJpsi as a function of the lepton helicity angle\footnote{The inclusion of charge-conjugate processes is implied throughout the paper.}. 
The measurement is based on a data sample corresponding to an integrated luminosity of $1.1 \invfb$ of proton-proton ($pp$) collisions collected with the \lhcb Upgrade~I detector in October 2024 at a centre-of-mass energy of 13.6\tev.

The \KJpsi decay mode provides a valuable benchmark for assessing the performance and systematic control of the upgraded \lhcb detector in many physics analyses. It is characterised by a large branching fraction and a precisely known angular distribution.  This is in contrast to \eg $B^0\to\jpsi K^{*0}$ decays, where interference with exotic contributions, such as \exoone decays, complicates a direct comparison of the angular distribution with Standard Model (SM) predictions. Tree-level \btosccbar decays, such as \KJpsi, also serve as crucial validation channels for several flagship LHCb analyses, such as those involving \decay{\Bu}{\Kp\mup\mun} and \decay{\Bd}{\Kp\pim\mup\mun} decays, which are mediated by the quark-level transition \btosmumu.
These transitions, along with their Cabibbo-suppressed \btodmumu counterparts,  are forbidden at tree level in the SM and are therefore highly sensitive to potential contributions from physics beyond the SM~\cite{Celis:2015ara, Falkowski:2015zwa, Crivellin:2015mga, Barbieri:2016las, Buttazzo:2017ixm}.
Over the past decade, a consistent pattern of deviations from SM predictions has emerged across various \btosmumu observables, generating significant attention and debate within the physics community. These tensions are present both in branching fraction measurements and angular analyses of these decay modes~\cite{LHCB-PAPER-2012-011, LHCB-PAPER-2016-012, LHCb-PAPER-2020-002, LHCB-PAPER-2014-006, LHCb-PAPER-2020-041, LHCb-PAPER-2021-014, LHCb-PAPER-2021-022, LHCB-PAPER-2015-009, ATLAS:2018gqc, CMS:2020oqb}. 
The data collected using the \lhcb Upgrade~I detector will help better characterise the size and dependencies of these tensions.
For these \btosmumu analyses, decays such as \KJpsi  will be used to calibrate the detector response, validate efficiency estimates, and cross-check the overall analysis strategy. This approach is motivated by the high statistical precision of these \btosccbar modes and the similar, or indeed identical, final states to their \btosmumu counterparts.

The normalised differential decay rate for \KJpsi, where the \Jpsi meson is reconstructed via the \Jpsimm decay,  is given as~\cite{Bobeth:2011nj}
\begin{equation}
\frac{1}{\Gamma}\frac{\mathrm{d}\Gamma}{\mathrm{d}\ctl} = \frac{3}{4}(1 - \fh)(1 - \cos^2\theta_{\ell}) + \frac{1}{2}\fh + \afb\ctl.
\label{eq:decayrate}
\end{equation}
Here, $\theta_\ell$ is defined as the angle between the $\mu^+$ ($\mu^-$) lepton and the direction opposite to that of the $B^+$ ($B^-$) meson in the dimuon rest frame. The angular observables \afb and \fh correspond to the forward-backward asymmetry and the flatness parameter, respectively. Since in the SM the region of dimuon mass squared (\qsq) studied in this work is dominated by the decay of the \jpsi, 
\ie a spin-1 vector meson, both of these observables are expected to be zero due to angular momentum conservation. 
In the SM, the values of these coefficients are also predicted to be zero in the corresponding \btosmumu mode, \Kmm, up to subleading effects~\cite{Bobeth:2011nj}.

The \ctl distribution is subject to distortions arising from detector acceptance, event selection, and reconstruction effects. Accurate modelling of these effects in simulated samples is essential for the reliable extraction of angular observables.
As shown in Eq.~\ref{eq:decayrate}, the differential decay rate for \KJpsi exhibits a dependence on \ctl up to second order. This is the same order to which both the hadron and lepton helicity angles 
are expanded in the angular distribution of 
\Kstmm and \Phimm  decays, where some of the most significant tensions with the SM are seen~\cite{LHCb-PAPER-2020-002, ATLAS:2018gqc, CMS:2024atz, LHCb-PAPER-2023-032, LHCb-PAPER-2024-011, LHCb-PAPER-2021-022}.  Having the same order means the angular variation should be at a similar scale and therefore sensitive to similar efficiency-effects. 

The forward-backward asymmetry, \afb, can be sensitive to detector- or reconstruction-induced asymmetries in \ctl. Conversely,  \fh is affected by symmetric distortions, such as mismodelled resolution or momentum-dependent efficiency effects that are independent of lepton charge. These effects can alter the relative contributions of the constant and quadratic terms in the angular distribution and thus bias \fh. 
Although the \KJpsi decay does not involve a hadronic helicity angle, the dominant experimental effects, particularly those related to tracking, reconstruction, and, to a large extent, candidate selection, are largely insensitive to the specific identities of the final-state particles. 
As a consequence, this work also helps validate the detector response for 
\btodmumu transitions.

In this work, the first analysis of a $b$-hadron decay using data collected with the \lhcb Upgrade~I detector is presented.
The angular observables \afb and \fh are measured separately for each magnetic field polarity, as well as using the combined dataset.
The polarity-separated measurements are further extracted differentially as functions of 17 kinematic and detector-response variables.
Each differential measurement is typically performed in ten equipopulated intervals~(bins) along a given variable. This strategy is designed to identify and mitigate potential trends that could introduce biases, particularly in \btosmumu measurements in the case where the underlying distributions of these variables differ between \btosccbar and \btosmumu decays. 

\section{Datasets and simulation}
\label{sec:Detector}

The \lhcb Upgrade I  detector~\cite{LHCb-DP-2022-002} is a 
single-arm forward spectrometer covering the pseudo\-rapidity range $2<\eta <5$, designed for 
the study of particles containing \bquark or \cquark quarks. 
The detector was installed prior to the Run 3 data-taking period, which started in 2022. 
The Upgrade I detector is a major change of the detector system and was designed to
match the performance of the Run 1--2 detector~\cite{LHCb-DP-2022-002,LHCb-DP-2008-001}, while allowing it to operate 
at approximately five times the instantaneous luminosity. 
The high-precision tracking system consists 
of a silicon-pixel vertex detector surrounding the $pp$
interaction region, a large-area silicon-strip detector located
upstream of a dipole magnet with a bending power of about
$4{\mathrm{\,T\,m}}$, and three stations of scintillating-fibre detectors downstream of the magnet.
The magnetic field deflects oppositely charged particles in opposite directions, which can lead to charge-asymmetric detection efficiencies. 
The configuration with the magnetic field pointing upwards is referred to as \MagUp, while the configuration with the field pointing downwards is referred to as \MagDown.
Different types of charged hadrons are distinguished using information
from two ring-imaging Cherenkov detectors, equipped with photon detection systems. 
Photons, electrons and hadrons are identified by a calorimeter system consisting of electromagnetic
and hadronic calorimeters. Muons are identified by a
system composed of alternating layers of iron and multiwire
proportional chambers.

The central feature of the upgraded detector is the readout of all subdetectors at the LHC crossing rate of 40\mhz, which allows the data to be processed in an all-software trigger~\cite{LHCb-TDR-016, LHCb-TDR-017, LHCb-TDR-018, LHCb-TDR-021}, thereby enabling the real-time reconstruction and selection of events.
The trigger system is implemented in two stages: the first-level trigger (\hltone), a GPU-based inclusive stage focused on charged-particle reconstruction that reduces the data volume by roughly a factor of 20; and the second-level trigger (\hlttwo), a CPU-based stage that performs the full offline-quality reconstruction and selection of physics signatures. 
A large disk buffer is placed between these stages to hold the data while the real-time alignment and calibration are being performed.
Triggered data further undergo a centralised, offline processing step~\cite{Sprucing}.
Subsequently, highly automated \lhcb analysis productions deliver
physics-analysis-ready data with very low latency across the entire \lhcb physics programme~\cite{Sprucing, FunTuple}.

The data used were recorded at the design instantaneous luminosity of \mbox{$\mathcal{L} = 2 \times 10^{33}~\mathrm{cm}^{-2}~\mathrm{s}^{-1}$}, corresponding to 5.3 visible $pp$ interactions per bunch crossing.   Of the total 1.1\invfb integrated luminosity analysed, 65\% was taken with the \MagDown configuration and 35\% in the \MagUp configuration. 

Simulation is required to model the effects on the \ctl distribution of the detector acceptance,  track reconstruction, and candidate selection requirements. 
In the simulation, $pp$ collisions are generated using
  \pythia~\cite{Sjostrand:2007gs,*Sjostrand:2006za} 
  with a specific \lhcb configuration~\cite{LHCb-PROC-2010-056}.
  Decays of unstable particles
  are described by \evtgen~\cite{Lange:2001uf}, in which final-state
  radiation is generated using \photos~\cite{davidson2015photos}.
  The interaction of the generated particles with the detector, and its response,
  are implemented using the \geant
  toolkit~\cite{Allison:2006ve, *Agostinelli:2002hh} as described in
  Ref.~\cite{LHCb-PROC-2011-006}.  The physics model used to generate \KJpsi candidates assumes that both  \afb and \fh are zero.

\section{Candidate selection and remaining backgrounds}
\label{sec:Selection}

Data are initially selected using the online trigger system. In \hltone, trajectories of charged particles that originate within the vertex detector acceptance and traverse the rest of the tracking system are identified. The momenta of these tracks are measured with percent-level precision, and they can be identified as either muons or electrons if they pass the relevant identification criteria~\cite{LHCb-DP-2022-002}. In order to pass the first-level trigger, either two tracks must form a common vertex that is significantly displaced from all primary vertices~(PVs) in the event, or a single track must be significantly displaced from all PVs. In the latter case, looser selection requirements are applied if the track is identified as a muon.

The \hlttwo stage uses information provided by the real-time alignment and calibration of the detector to perform an offline-quality reconstruction~\cite{LHCb-DP-2022-002,LHCb-PROC-2015-011}. This reconstruction is then used to apply further, more stringent, requirements: all tracks must have significant impact parameter~(IP) values with respect to all PVs and a low track-fit $\chi^2$, as determined from a Kalman fitter;  two oppositely charged muon tracks and an additional track should form a common vertex that is displaced from every PV;  the resulting \Bu candidate must have a small IP with respect to its associated PV (taken as the PV that best fits the \Bu flight direction); and muon tracks should have appropriate signatures in both the muon and RICH detectors.  No explicit particle identification~(PID) requirement is applied to the accompanying hadron during the online selection procedure.  
Tracks are removed if their hit pattern indicates there is a high probability they arose from incorrectly combining hits belonging to different particles.

At the offline stage, the IP requirements from the trigger are further tightened. Minimum requirements on the angles between final-state particle trajectories are applied to ensure that the \Bu candidates are not constructed from duplicate tracks that reuse the same segment in the vertex detector. 
Additional PID criteria are also applied to each of the final-state particles.  For this, the output of artificial neural networks trained to identify each charged-particle species $i$, normalised between 0 and 1 and named $P^\text{NN}_i$~\cite{LHCb-DP-2014-002,LHCb-DP-2018-001}, is employed. 
The combined requirements on the $P^\text{NN}_i$ outputs correspond to a signal efficiency of approximately 84\%, and are purposefully chosen to achieve signal and background rejection efficiencies similar to those of the PID criteria used to select \btosmumu processes during Run 2 data taking.

It is required that the reconstructed mass of the \Bu candidates,  denoted \mKmm,  lies in the range $5200 < \mKmm < 5850$\mevcc and that the muon pair has a mass within $\pm$100\mevcc of the known $\Jpsi$ mass value~\cite{PDG2024}. 
Contributions from combinatorial backgrounds, formed by incorrect vertexing of tracks not originating from the same \mbox{$b$-hadron} decay, can be reduced using a gradient-boosted decision-tree~(BDT) classifier~\cite{CatBoost}. 
Given the high purity of the \KJpsi mode, employing a  BDT in the selection does not result in a large improvement on the signal significance. Nevertheless, a loose selection is applied on the BDT output, as multivariate classifiers form an integral part of the selection for rare \btosmumu transitions, for which this work serves as a benchmark. This BDT classifier 
is trained using topological and kinematic variables. 
For the signal proxy, \KJpsi candidates in the range $5200 < \mKmm < 5500$\mevcc are used, with residual background contributions subtracted using the \sPlot technique~\cite{Pivk:2004ty} and \mKmm taken as the discriminating variable. The background proxy is taken from \KJpsi candidates with $\mKmm > 5500$\mevcc.
To ensure independence between the events used to train the BDT and those to which the BDT is applied, cross-validation in the form of $k$-folding~\cite{kFold} ($k = 10$) is used, \ie, ten BDTs are trained, each using a different nine-tenths of the data.  
The corresponding BDT efficiency on signal is around  99\%, and a cross-check is performed by tightening this selection, as discussed later in Sec.~\ref{sec:results}. 

After all selection criteria are applied, the only significant backgrounds are either combinatorial in nature or those arising from Cabibbo-suppressed \PiJpsi decays where the pion is misidentified as a kaon. The contribution from the latter is below 1\% of the signal yield. Both these backgrounds are accounted for in the fit model.
Of the \KJpsi candidates considered in the fit, 0.8\%  originate from the same bunch crossing as another \KJpsi candidate and are referred to as multiple candidates.  In the baseline selection, one candidate per event is arbitrarily selected.

As outlined in the introduction, some fits are performed on subsets of the data to better validate the understanding of the detector response.  
Data are split according to magnet polarity, and then further subdivided according to the number of primary vertices ($N_{\mathrm{PV}}$); the number of tracks with hits in all tracking subdetectors ($N_{\mathrm{Tk}}$); the transverse momentum and pseudorapidity of the $\Bu$, $\jpsi$, and kaon candidates; the impact parameter $\chi^2$ ($\chi^2_{\mathrm{IP}}$), which quantifies the degradation of the PV fit when including the track of interest, of the $\jpsi$ and kaon candidates; the vertex fit quality ($\chi^2_{\mathrm{vtx}}$) of the $\Bu$  mesons; the $\chi^2$ of the \Bu flight distance ($\chi^2_{\mathrm{FD}}$), defined as the separation between the $\Bu$ decay vertex (DV) and the PV; the $\Bu$ momentum and its angle with respect to the PV--DV vector ($\cos\theta_{\mathrm{DIRA}}$); the maximum momentum of the $\jpsi$ decay products ($\max(p(\jpsi))$); the pseudorapidity of the muon with the same charge as the $B$ meson ($\eta(\mu^\pm)$);  and the opening angle between the two muons, $\theta(\mu^+\mu^-)$. Each variable is used to divide the data into ten approximately equipopulated bins, with the exception of $N_{\mathrm{PV}}$, where six bins are used. 
These variables are chosen either to test the stability of the reconstruction under different conditions (\eg, varying occupancies or decay-vertex quality), to ensure that the efficiency function correctly reproduces strong correlations with the variable in question (as is the case for the kinematic variables), or because their distributions differ significantly between \btosccbar and \btosmumu transitions (as in the case of $\theta(\mu^+\mu^-)$ and $p_T$ of the $J/\psi$ meson).

\section{Calibration of simulation and determination of the efficiency model}
\label{sec:Sim}
Simulated events must be calibrated to accurately reflect all relevant features of the LHC production environment and the performance of the \lhcb detector. This calibration is implemented through a set of weights, whose product is applied to simulation to correct for differences in particle identification, trigger efficiencies, \Bu production kinematics, detector occupancy, and track reconstruction.

The PID calibration weight, $w_{\text{PID}}$, is derived as the ratio of PID efficiencies in data and simulation. The efficiency is evaluated as a function of the track momentum and pseudorapidity, using calibration samples composed of muon candidates from \Jpsimm decays and kaon candidates from \Dzpim decays~\cite{LHCb-PUB-2016-021}.\footnote{Only $\Jpsi$ decays characterised by a significant flight distance with respect to the associated PV are considered for calibration.} 

The efficiency for a \KJpsi candidate to pass the \hltone requirements used in this work is calibrated as a function of the \Bu transverse momentum using the \textsc{TriggerCalib} package~\cite{Albrecht:2025une}.  
To calculate the efficiency in data, the denominator is constructed from \KJpsi candidates in events where tracks not associated with said candidate have passed an alternative set of \hltone selections.
The efficiency is then given by the fraction of these events where the \KJpsi candidate passed one of the \hltone selections of interest~\cite{LHCb-PUB-2014-039}.  
This efficiency is deduced from both data and simulation,  and the correction weight $w_{\rm \hltone}$ is obtained from their ratio. 
In data, the signal yield is extracted by subtracting the background extrapolated from the \mKmm sidebands.

To correct for the remaining mismodelling of the detector occupancy,  \Bu meson production kinematics, and the quality of the reconstruction, a weight \wkin is applied. The detector occupancy is corrected using $N_{\mathrm{PV}}$ and $N_{\mathrm{Tk}}$;  the \Bu meson production kinematics are corrected using its momentum $p$ and transverse momentum \pt; and the reconstruction quality is corrected using the $\chi^2_{\mathrm{IP}}$ and $\chi^2_{\mathrm{vtx}}$ of the \Bu meson. 
The weighting is performed using \KJpsi data candidates as the target, with background contributions subtracted using the \sPlot technique~\cite{Pivk:2004ty}, where the mass \mKmm is used as the discriminating variable. To ensure statistical independence between the calibration and measurement samples, the $k$-folding technique~\cite{kFold} is employed. The binning scheme used in the weighting is optimised using decision trees trained with the \texttt{GBReweighter} algorithm~\cite{Rogozhnikov:2016bdp}.

\begin{figure}[tb]
    \centering
    \includegraphics[width=0.49\textwidth]{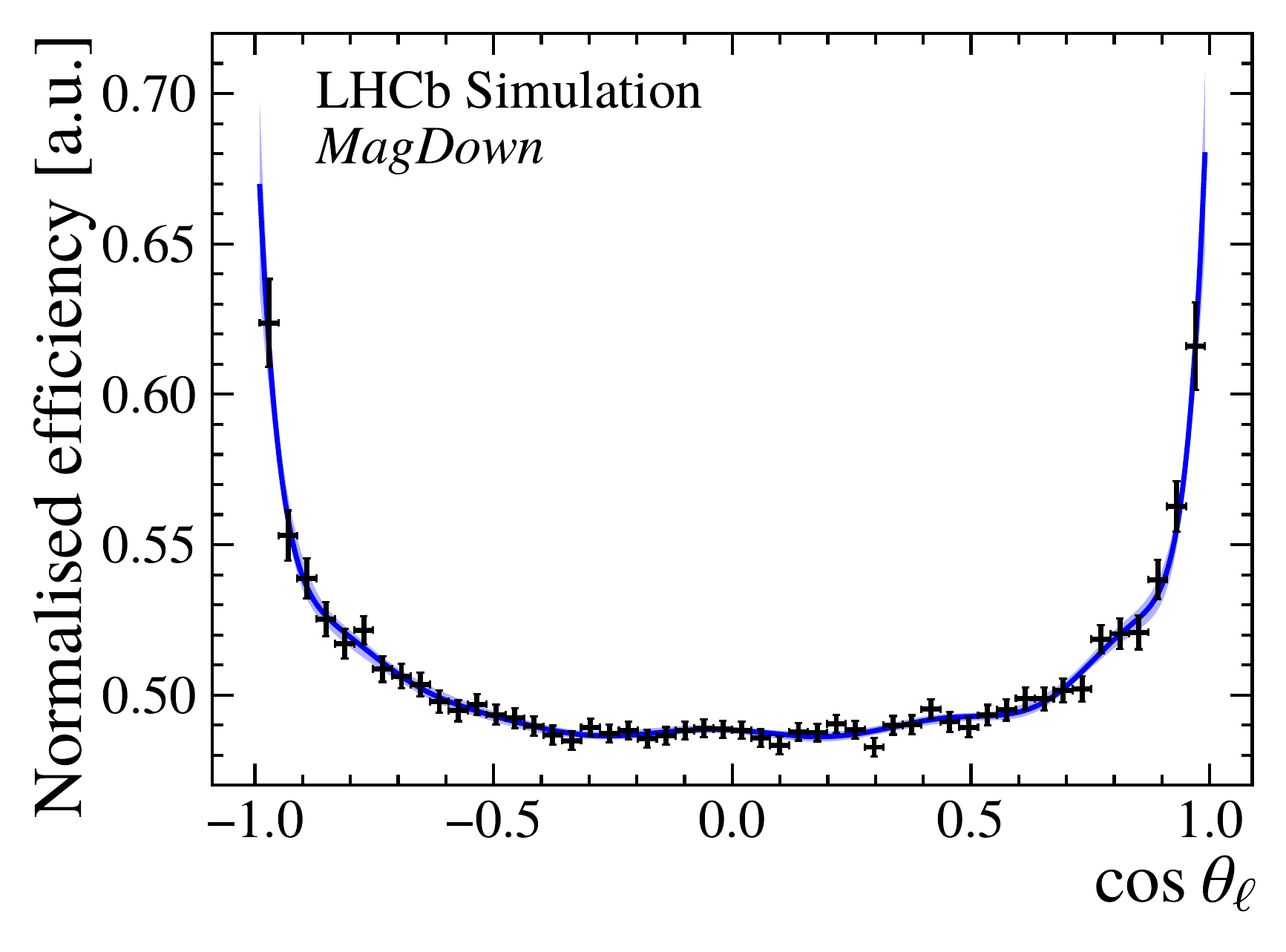}    
    \includegraphics[width=0.49\textwidth]{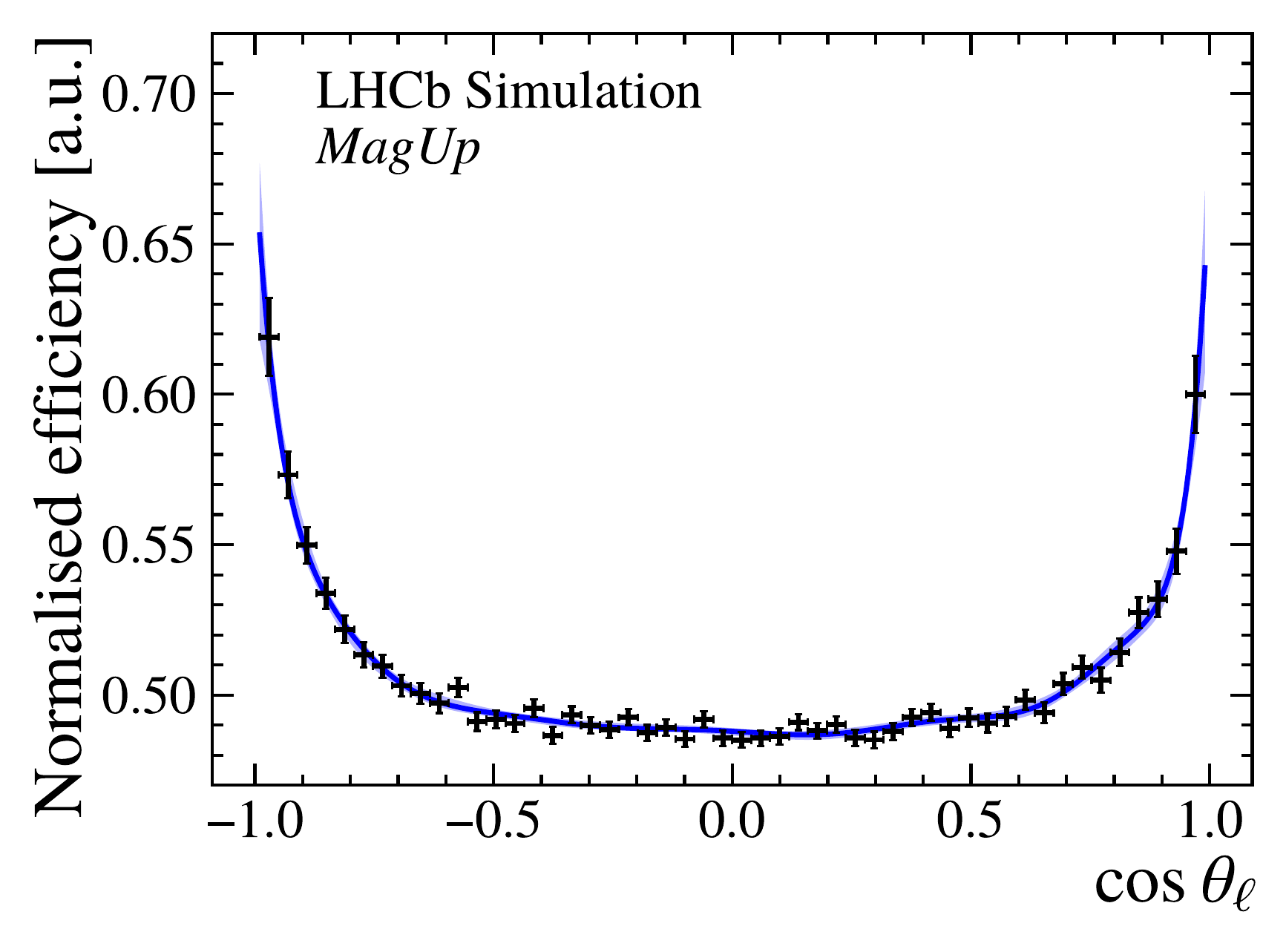}  
    \caption{ Normalised efficiencies as a function of $\cos{\theta_\ell}$ for (left)~\emph{MagDown} and (right)~\emph{MagUp} magnet polarities. The efficiency parameterisation, shown as the blue curve, is obtained using the method of moments with a Legendre-polynomial expansion truncated at order 12. The shaded blue band represents the 68.27\% confidence interval derived from the uncertainties on the polynomial coefficients. The black points indicate the distribution of the simulated $\decay{B^+}{K^+ \mu^+ \mu^-}$ decays used to determine the efficiency function.
    }
    \label{fig:efficiencies}
\end{figure}

To obtain the efficiency as a function of \ctl, $\epsilon(\ctl)$, the correction weights are applied to simulated samples of \KJpsi candidates with all selections applied.   The efficiency distribution is obtained by weighting the simulation by the inverse of $\mathrm{d}\Gamma/\mathrm{d}\ctl$, and is parameterised using a 12th-order Legendre polynomial. 
Much of the resulting shape is determined by the detector’s geometrical acceptance and the momentum requirements imposed on the final-state particles.
The \ctl range is restricted to \mbox{$-0.99 < \ctl < 0.99$}, as values beyond this lead to undesirably large weights due to the divergence of the normalisation factor \mbox{$(\mathrm{d}\Gamma/\mathrm{d}\!\ctl)^{-1} \propto 1/(1 - \cos^2\theta_l)$}.
 The parameterisation is performed separately for the \MagUp and \MagDown simulated samples, as shown in Fig.~\ref{fig:efficiencies}. 
 The goodness-of-fit of this parameterisation is assessed by verifying that fitting the reconstructed simulated \ctl distribution with Eq.~\ref{eq:decayrate}, multiplied by the parameterised efficiency $\epsilon(\ctl)$, yields values of \fh and \afb consistent with those used to generate the samples. 
Separate efficiency functions are parameterised each time a fit is performed to a subset of the data, as is the case when evaluating the trend of the angular coefficients across the aforementioned kinematic and topological variables, or when performing other cross-checks discussed in Sec.~\ref{sec:results}.

\section{Fit model}
\label{sec:FitModel}

To better discriminate \KJpsi candidates from combinatorial and \PiJpsi backgrounds, a two-dimensional unbinned extended maximum-likelihood fit is performed to the \mKmm and \ctl distributions using the \zfit software~\cite{jonas_eschle_2019_2602043,Eschle:2019jmu}. The two distributions are assumed to factorise, which is verified by inspecting the \ctl distribution in different regions of \mKmm, and computing both Pearson’s correlation coefficient~\cite{Pearson1895} and the Hilbert–Schmidt independence criterion~\cite{HSIC}.

The \ctl distribution of the \KJpsi candidates is modelled using the 
normalised differential decay rate given by Eq.~\ref{eq:decayrate}, multiplied by the efficiency function, $\epsilon(\ctl)$. The \mKmm distribution is modelled using a superposition of two Gaussian functions and a double-sided Crystal Ball (CB) function, where the latter consists of a Gaussian distribution core with power-law tails on both the upper-mass and lower-mass sides~\cite{Skwarnicki:1986xj}. All three functions share the same peak position. The parameters describing the mass shape are fixed from fits to simulated samples. However, to account for residual data-simulation differences,  two additional fit parameters are introduced in the fit to data. One represents a shift in the peak position of the composite distribution, and the other a common scaling factor on the width of the distribution. 

For \PiJpsi background candidates, the \ctl distribution is modelled with second-order Chebyshev polynomials, while the \mKmm distribution is described by a Gaussian function combined with a double-sided CB function. All parameters are fixed from simulation, except for the shift and scale factors of the \mKmm distribution, which are shared with the signal.
The relative number of \PiJpsi to \KJpsi candidates is left freely varying in the fit, and is found to be at the level of 0.3--0.5\%.  This is in good agreement with expectations based on the relative branching fractions and selection efficiencies. 

Finally, the combinatorial background is modelled with a fourth-order Chebyshev polynomial in \ctl and an exponential function in \mKmm. For the combinatorial component, all parameters are free to vary in the fit.

Pseudoexperiments generated with properties representative of the data show that the fit is unbiased and that confidence intervals obtained using the resulting covariance matrix provide correct coverage. This is true for both the integrated fit and for fits performed to subsets of the data. 

\section{Systematic uncertainties }
\label{sec:syst}
Systematic uncertainties on \afb and \fh are estimated using ensembles of pseudoexperiments, generated with variations applied either to the efficiency function, $\epsilon(\ctl)$, or to the fit model. 
All systematic sources are assessed by fitting pseudoexperiments twice: once with the baseline configuration and once with the varied configuration. The sum in quadrature of the mean and width of the difference in the fit results across the ensemble of pseudoexperiments is taken as the systematic uncertainty. Due to potential differences in detector response, the systematic uncertainties are evaluated separately for \MagUp and \MagDown data.  
A full summary of the included systematic sources is given in Table~\ref{tab:syst}.

Two dominant sources of systematic uncertainty are identified. The first arises from the limited size of the 
simulation samples used in the efficiency parameterisation. To assess this uncertainty, the simulated samples are bootstrapped with replacement~\cite{efron:1979}. For each bootstrap iteration, a new efficiency function is parameterised and used to generate a pseudoexperiment. This pseudoexperiment is then fitted using the corresponding variation of the acceptance function, and the resulting spread across the pseudoexperiments is taken as the systematic uncertainty. In the \MagUp (\MagDown) dataset, this uncertainty is at the level of 70\% (50\%) of the statistical uncertainty for both \afb and \fh. 

The second dominant source arises from the application of an alternative weighting scheme for obtaining the \wkin weights. In this alternative scheme, the momentum and transverse momentum of the $\jpsi$, rather than the $B^+$, are used. Additionally, events are divided depending on whether at least one final-state particle has a momentum above $100\gevc$, and the \wkin weights for these two independent datasets are derived separately.  
This threshold is motivated by the fact that high-momentum tracks are associated with additional challenges, such as poorer DV reconstruction quality and passage through detector regions with higher occupancy. When applying the baseline strategy, small differences in the angular coefficients are observed between these two event categories. These differences are mitigated by the alternative weighting scheme, and the resulting impact on the baseline measurement is therefore assigned as a systematic uncertainty using the approach outlined above. For \fh, this contribution is at the level of 70\% (50\%) of the statistical uncertainty for the \MagUp (\MagDown) dataset, while for \afb it remains below 10\%.

The majority of the remaining systematic sources contribute less than 20\% of the statistical uncertainty. 
The considered configurations, listed in order of decreasing impact on the resulting uncertainty, are: variations to the corrections applied to simulation; modifications to the truth-matching criteria in simulation; changes to the signal mass model; alterations to the hyperparameters used in the kinematic and multiplicity corrections; treatment of multiple candidates from the analysed dataset; variation of the polynomial order used to parameterise the combinatorial \ctl model; residual miscalibration in the \hlttwo trigger; uncertainty associated with particle identification corrections; and variation of the polynomial order used to parameterise the $\epsilon(\ctl)$ function.

The robustness of the results against the applied simulation corrections is assessed by producing an ensemble of alternative $\epsilon(\ctl)$ functions constructed removing all correction weights $w_{\rm PID}$,  $w_{\rm \hltone}$, and \wkin.
In simulation, a truth-matching algorithm is used to identify signal decays. The robustness of the signal definition is evaluated using stricter truth-matching criteria, assessing the impact of this choice by using the alternative efficiency function obtained.
For the signal mass model, two double-sided CB functions are used instead of the baseline combination of one double-sided CB and two Gaussian functions. The parameters are taken from simulation, with the scaling parameter derived from data. 
Variations in the hyperparameters used in the kinematic and multiplicity correction procedure are also considered, effectively changing the binning scheme used to obtain the \wkin weights.
To assess the impact of multiple candidates, the analysis is rerun without their removal.
The effect of the background model is evaluated by increasing the order of the Chebyshev polynomial used for the combinatorial background in \ctl from four to six. 
Sensitivity to any residual miscalibration in the efficiency of \hlttwo tracking is assessed by introducing weights representing the ratio of tracking efficiencies in data and simulation. These weights vary linearly across the pseudorapidity of the final-state tracks, 
and are obtained separately for each detector quadrant.
This serves to capture differences arising from imperfections in the simulation of the tracking system that are not currently modelled.
A new $\epsilon(\ctl)$ function is derived using these weights, and the resulting difference is taken as a systematic uncertainty.
The impact of the uncertainties due to the finite size of the data samples used  to obtain the particle identification corrections is assessed 
via an ensemble of alternative efficiency functions derived with correction weights varied within their uncertainties.
Finally, sensitivity to the functional form of the efficiency parameterisation is evaluated by increasing the polynomial order from 12 to 14.

Additional systematic variations are investigated but found to 
be negligibly small.
These include: modifying the set of selections used to derive $w_{\rm \hltone}$; 
calculating $w_{\rm \hltone}$ in two dimensions across both $p$ and \pt of the \Bu candidate; deriving the PID calibration as a function of $N_{\mathrm{Tk}}$; varying the angular resolution obtained on simulation and varying the requirement used to remove tracks that may share hits.

\begin{table}[bt]
\centering
\caption{Absolute systematic uncertainties for data taken in either the \MagDown or \MagUp configuration  for the  \afb and \fh observables. All numbers are in units of $10^{-3}$.}
\begin{tabular}{l|rr|rr}
\hline

Systematic source & \multicolumn{2}{c|}{\MagDown} & \multicolumn{2}{c}{\MagUp} \\
                           & \afb & \fh & \afb & \fh \\
\hline
Simulation sample size & 0.44 & 1.03 & 0.39 & 1.04 \\
Alternative scheme for \wkin & 0.03 & 1.14 & 0.04 & 0.90 \\
Simulation corrections & 0.14 & 0.30 & 0.12 & 0.24 \\
Truth-matching & 0.07 & 0.28  & 0.02 & 0.51 \\
Signal mass model & 0.01 & 0.25 & 0.01 & 0.17 \\
Choice of hyperparameters in \wkin & 0.01 & 0.19 & 0.01 & 0.17 \\
Multiple candidates removal & $<0.01$ & 0.05 & 0.07 & 0.20 \\
Orders of combinatorial & 0.04 & 0.08 & 0.04 & 0.10 \\
Tracking corrections & 0.01 & 0.06 & 0.01 & 0.07 \\
Uncertainty on PID corrections & 0.01 & 0.02 & $<0.01$ & 0.01 \\
Higher order efficiency & $<0.01$ & 0.01 & 0.01 & 0.01 \\
\hline
Total systematic uncertainty & 0.47 & 1.62 & 0.42 & 1.53 \\

Total statistical uncertainty & 0.59 & 1.41 & 0.80 & 1.91 \\

\hline
\end{tabular}
\label{tab:syst}
\end{table}

\section{Results and cross-checks}
\label{sec:results}
\begin{figure}[tb]
    \centering
    \includegraphics[width=0.49\textwidth]{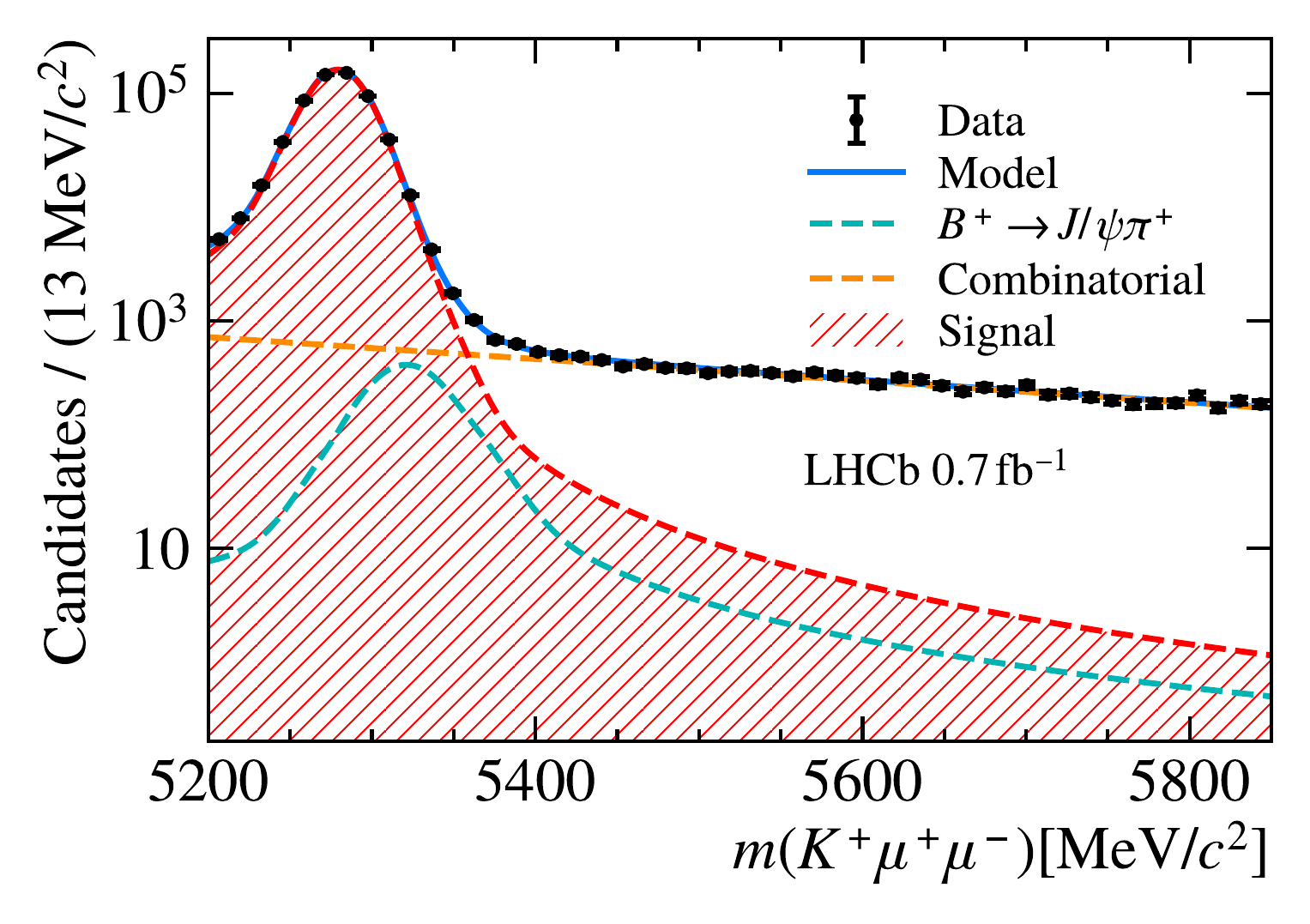}
    \includegraphics[width=0.49\textwidth]{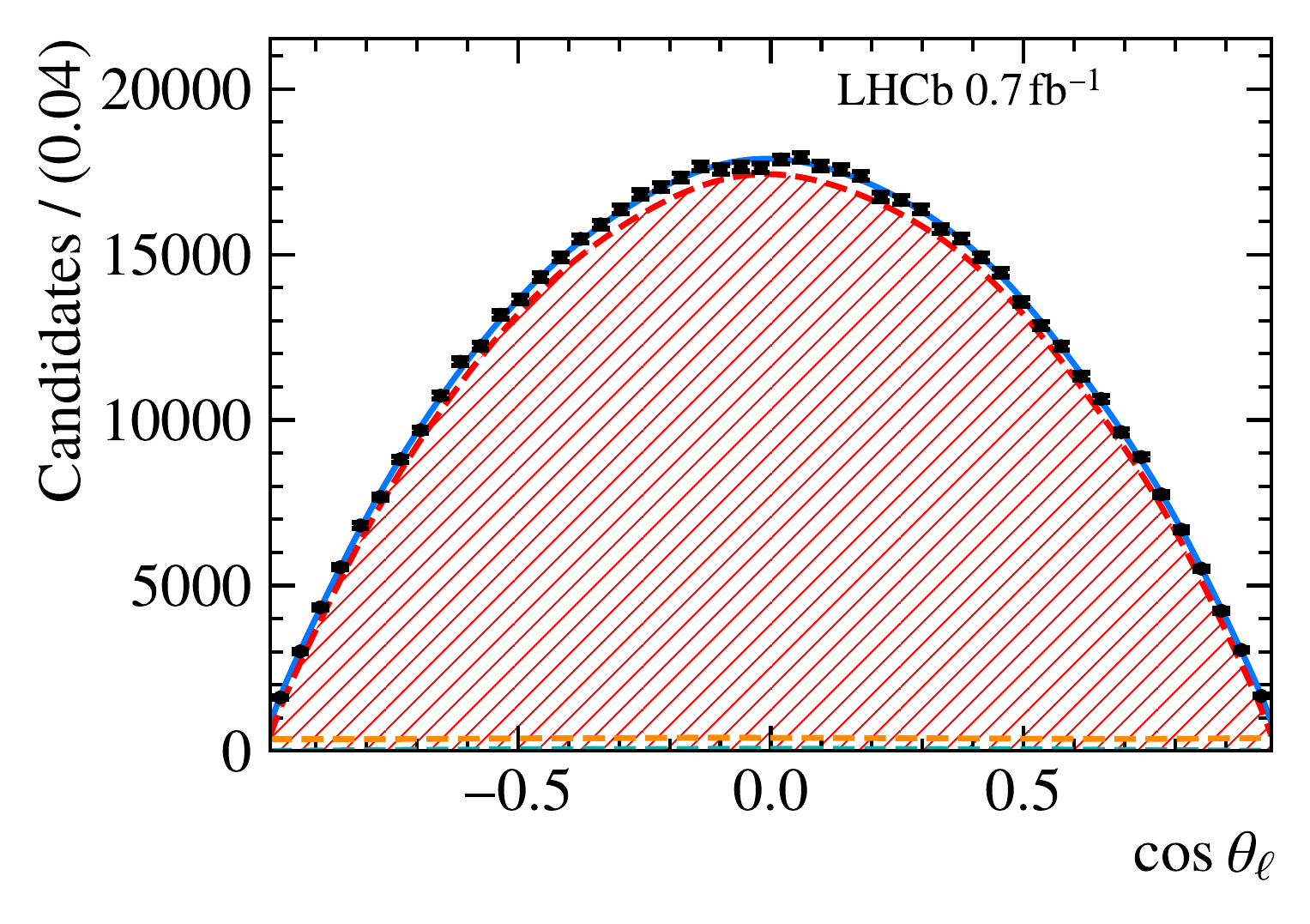}\\ 
    \includegraphics[width=0.49\textwidth]{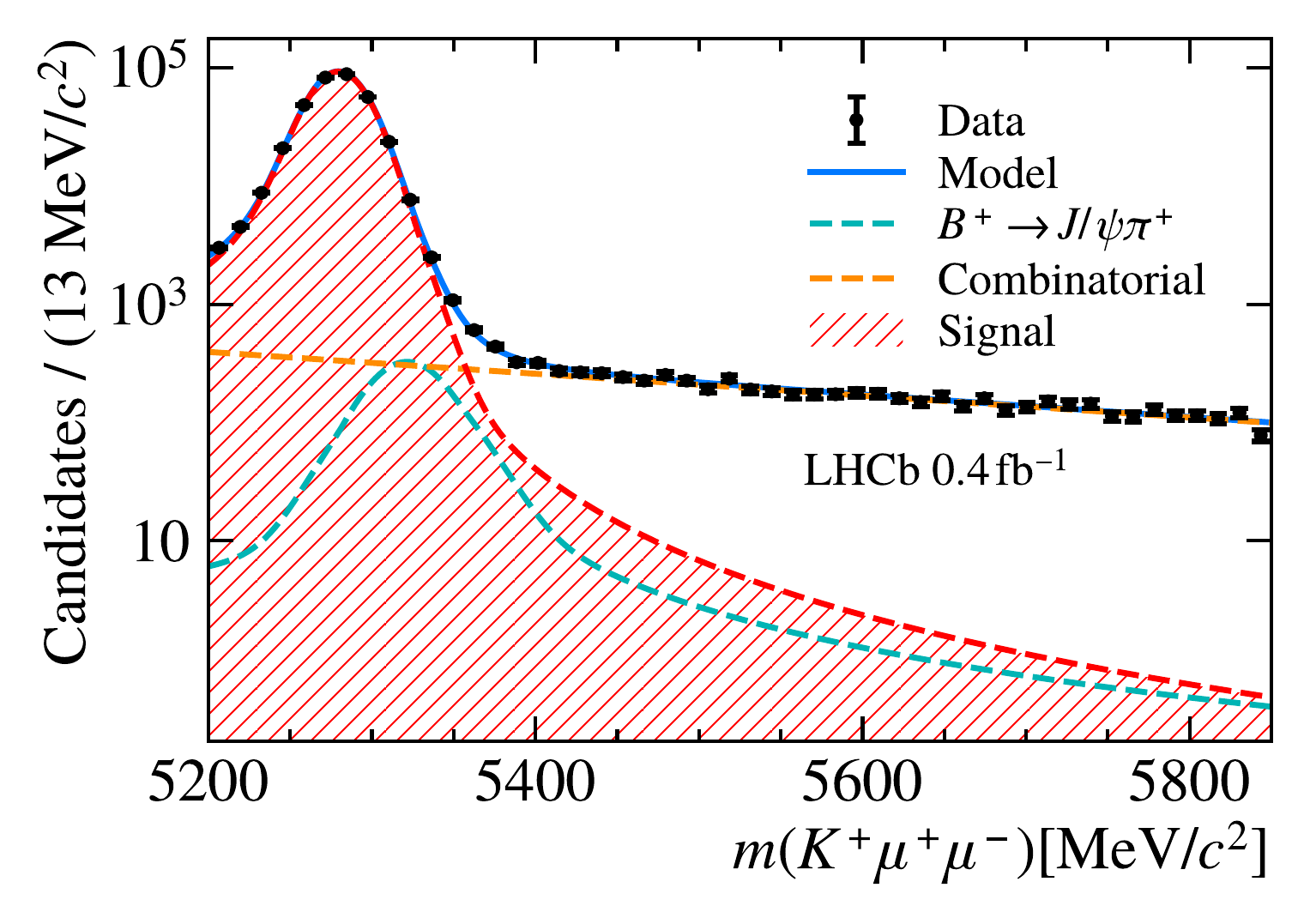}
    \includegraphics[width=0.49\textwidth]{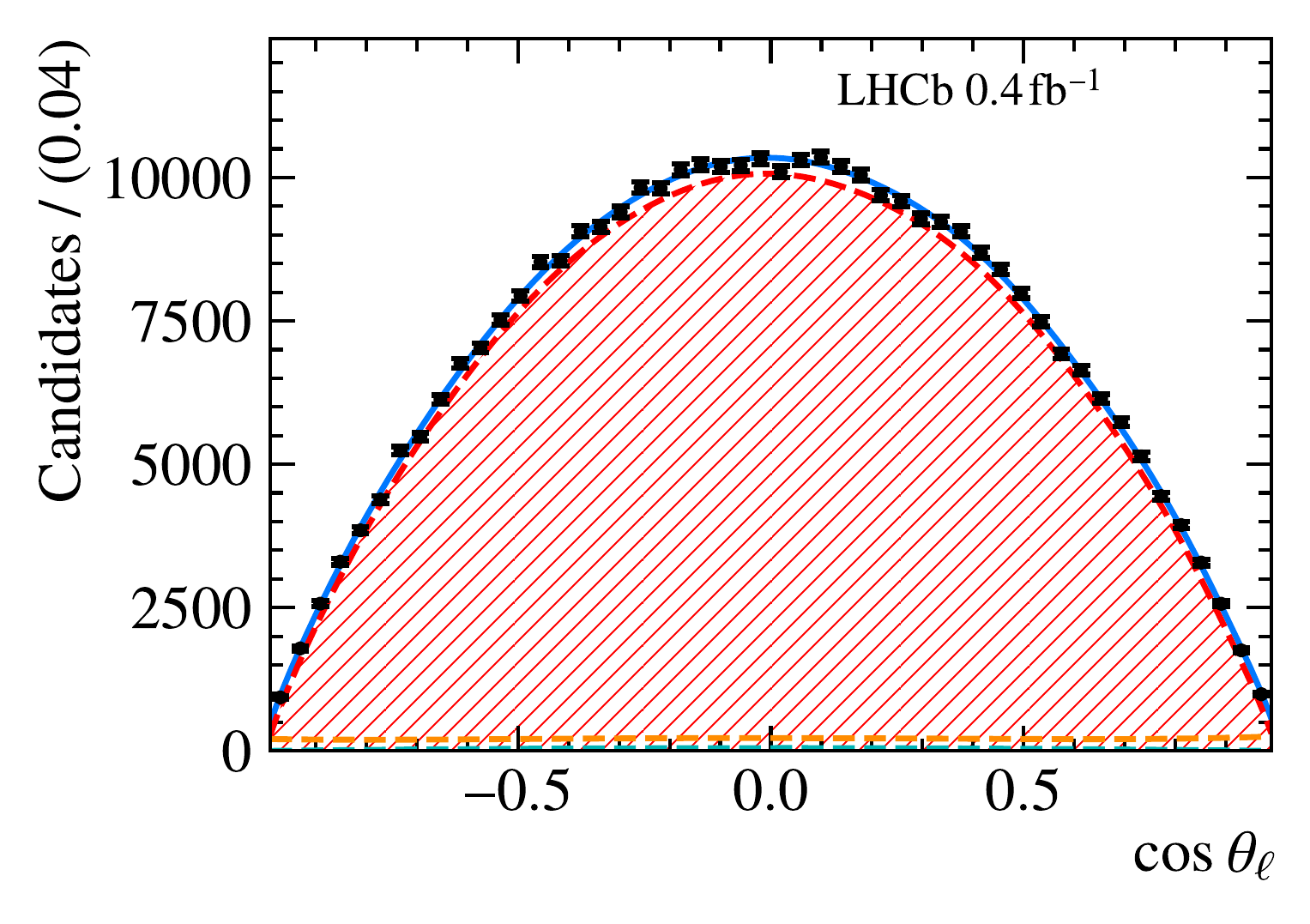}\\ 
    \caption{Projections of the fits onto $m(K^+\mu^+\mu^-)$ and $\cos{\theta_\ell}$ for  (top) \emph{MagDown}  and  (bottom) \emph{MagUp} data.}
        \label{fig:FitProjections}
\end{figure}

The projections of the fits to the \mKmm and \ctl distributions for \MagUp and \MagDown data are shown in Fig.~\ref{fig:FitProjections}, with results presented in Table~\ref{tab:AFBFHResults}. The average over both \MagDown and \MagUp is computed by weighting according to the inverse of the covariance matrix of each polarity. All systematic uncertainties are assumed to be 100\% correlated, except for the dominant systematic from the limited simulation size, which is treated as uncorrelated between the two polarities.
As can be seen from Table~\ref{tab:AFBFHResults}, all results agree with their predictions at the level of 1.2 standard deviations ($\sigma$).
The results obtained for the two magnet polarities agree with each other at the level of 1.5$\sigma$.
The pair-wise Pearson correlation between \fh and \afb is at the level of 5\%. This small correlation is expected due to the orthogonality of the angular functions associated with these coefficients.

\begin{table}[tb]
\centering
\caption{Measured values of \afb and \fh for \MagDown and \MagUp data, and their average. The first uncertainty is statistical, and the second systematic. All numbers are in units of $10^{-3}$.}\label{tab:AFBFHResults}
    \begin{tabular}{lcc}
        \hline
        & \afb & \fh \\
        \hline
        \MagDown & $-0.27 \pm 0.59\stat \pm 0.47\syst$ & $-1.2 \pm 1.4\stat \pm 1.6\syst$ \\
        \MagUp   & $\phantom{-}0.86 \pm 0.80\stat \pm 0.42\syst$ &  $\phantom{-}3.1 \pm 1.9\stat \pm 1.5\syst$ \\
        Average  & $\phantom{-}0.19 \pm 0.48\stat \pm 0.33\syst$ &  $\phantom{-}0.5 \pm 1.1\stat \pm 1.4\syst$\\
        
        \hline
    \end{tabular}
\end{table}

Further cross-checks are performed by varying the selection criteria and assessing the agreement with the baseline result, taking into account the statistical overlap. These checks are performed separately for the \MagUp and \MagDown datasets.  

Much tighter PID requirements with respect to the baseline selection are applied, namely $P^\text{NN}_K > 0.2$  is changed to $P^\text{NN}_K > 0.6$ and $P^\text{NN}_{\mu} > 0.2$ is changed to $P^\text{NN}_{\mu} > 0.8$ for the kaon and muon candidates, respectively.
All observed shifts are compatible with the baseline result at the level of 2$\sigma$ or less.

To replicate a BDT selection more representative of that used in analyses of \btosmumu decays, the BDT requirement is tightened, resulting in a signal efficiency of 65\%. 
All observed shifts in \afb and \fh are compatible with the baseline result at the 1$\sigma$ level or below, except for \afb in the \MagUp polarity, which is compatible at 2.3$\sigma$. 

\begin{figure}[tb]
    \centering
    \includegraphics[width =1\textwidth]{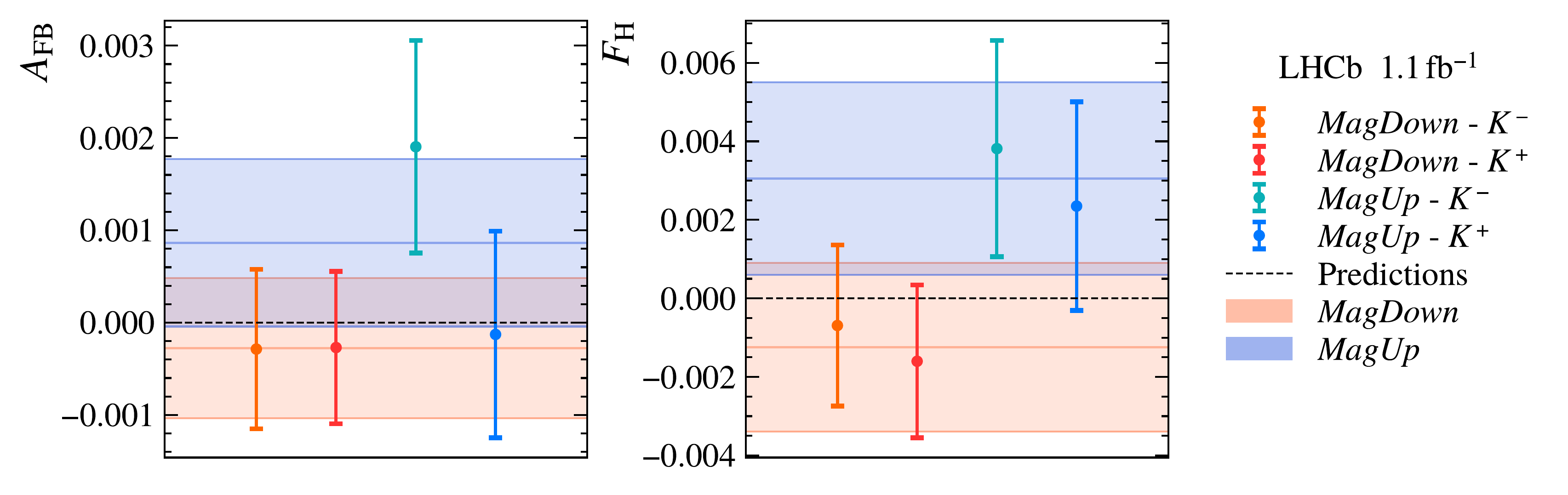}

    \caption{
    Measured values of $A_{\rm FB}$ and $F_{\rm H}$ obtained by splitting the dataset according to the kaon charge, including only statistical uncertainties. The horizontal bands indicate the corresponding values obtained without the charge split, with both statistical and systematic uncertainties included.}
    \label{fig:kaoncharge}
\end{figure}

The results are also obtained by splitting the data according to the charge of the kaon, and are displayed in Fig.~\ref{fig:kaoncharge}. The uncertainties shown are purely statistical, and no significant pattern or deviation is observed. Similar behaviour is exhibited when data are split either according to which quadrant of the muon station the muon with the same charge as the $B$ meson entered, or which quadrant of the scintillating-fibre detectors the kaon entered. 
The results are also checked when splitting according to whether at least one of the final-state tracks has a momentum larger than $100\gevc$, as discussed in Sec.~\ref{sec:syst}. Finally, to assess the stability of the results against variations in data-taking conditions, the analysis strategy is applied to an independent sample of data collected earlier in 2024, recorded using a different trigger configuration and under lower instantaneous luminosity conditions. The extracted angular coefficients from this sample are consistent with the baseline results.

\begin{figure}[tb]
    \centering
    \includegraphics[width=0.95\textwidth]{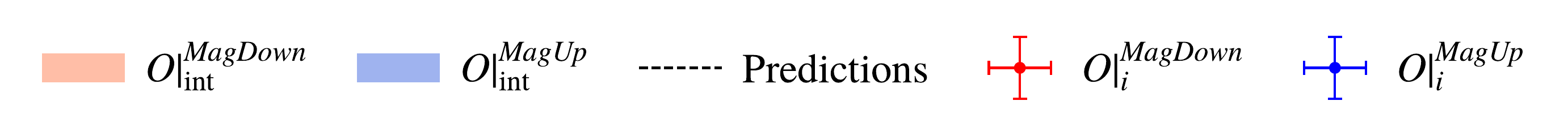}\\ 
    \includegraphics[width=0.45\textwidth]{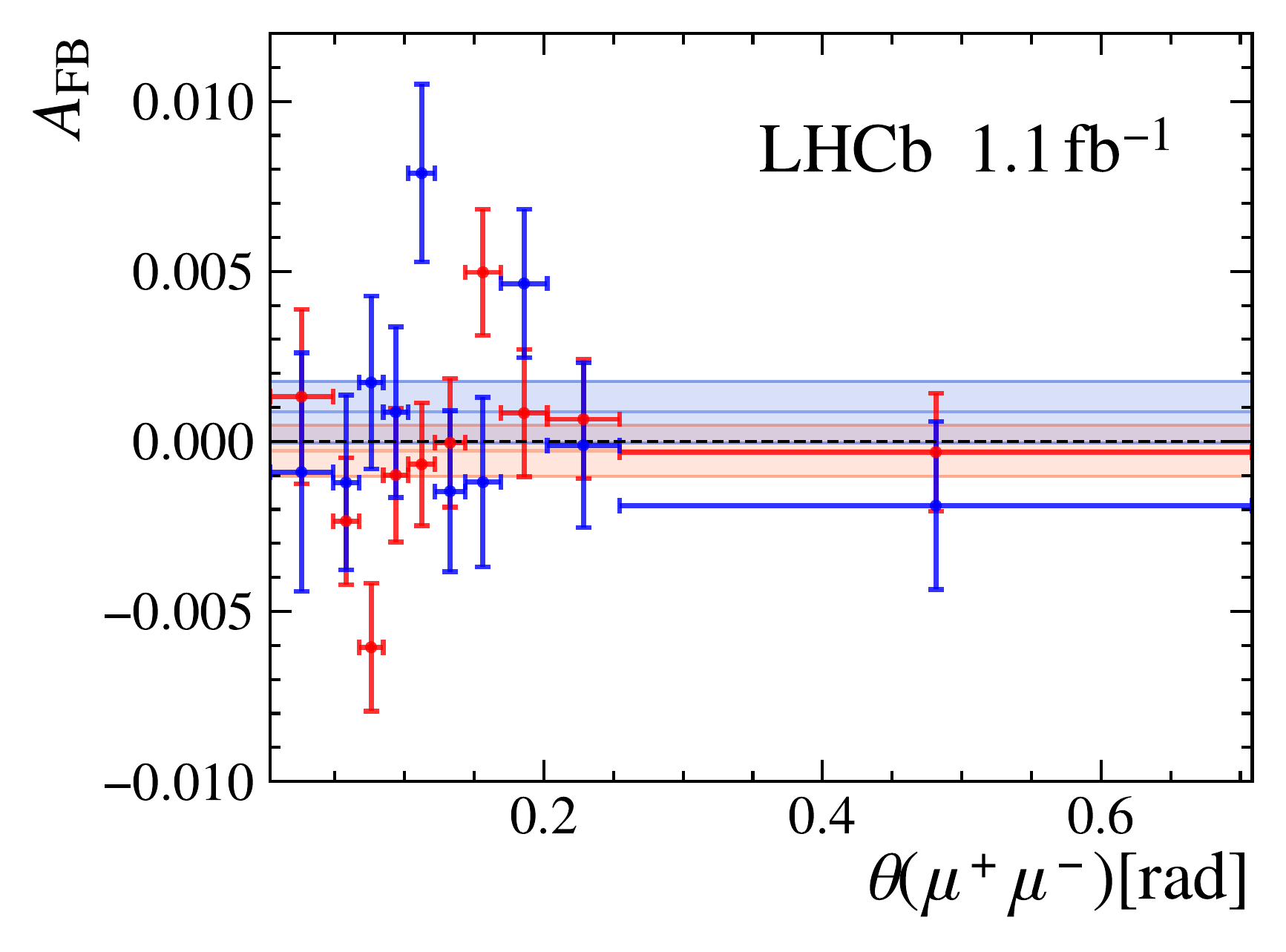}
    \includegraphics[width=0.45\textwidth]{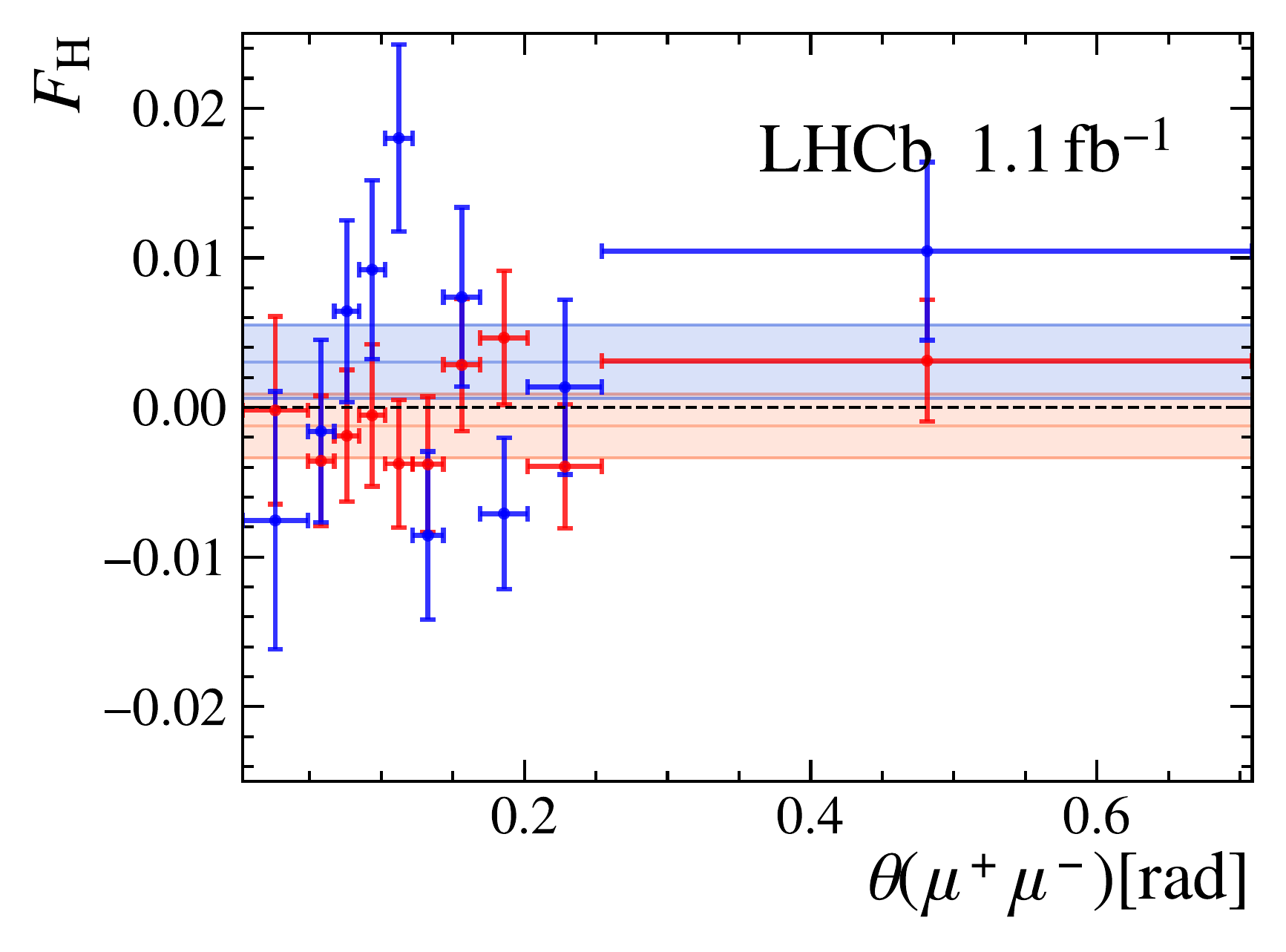}\\ 
    \includegraphics[width=0.45\textwidth]{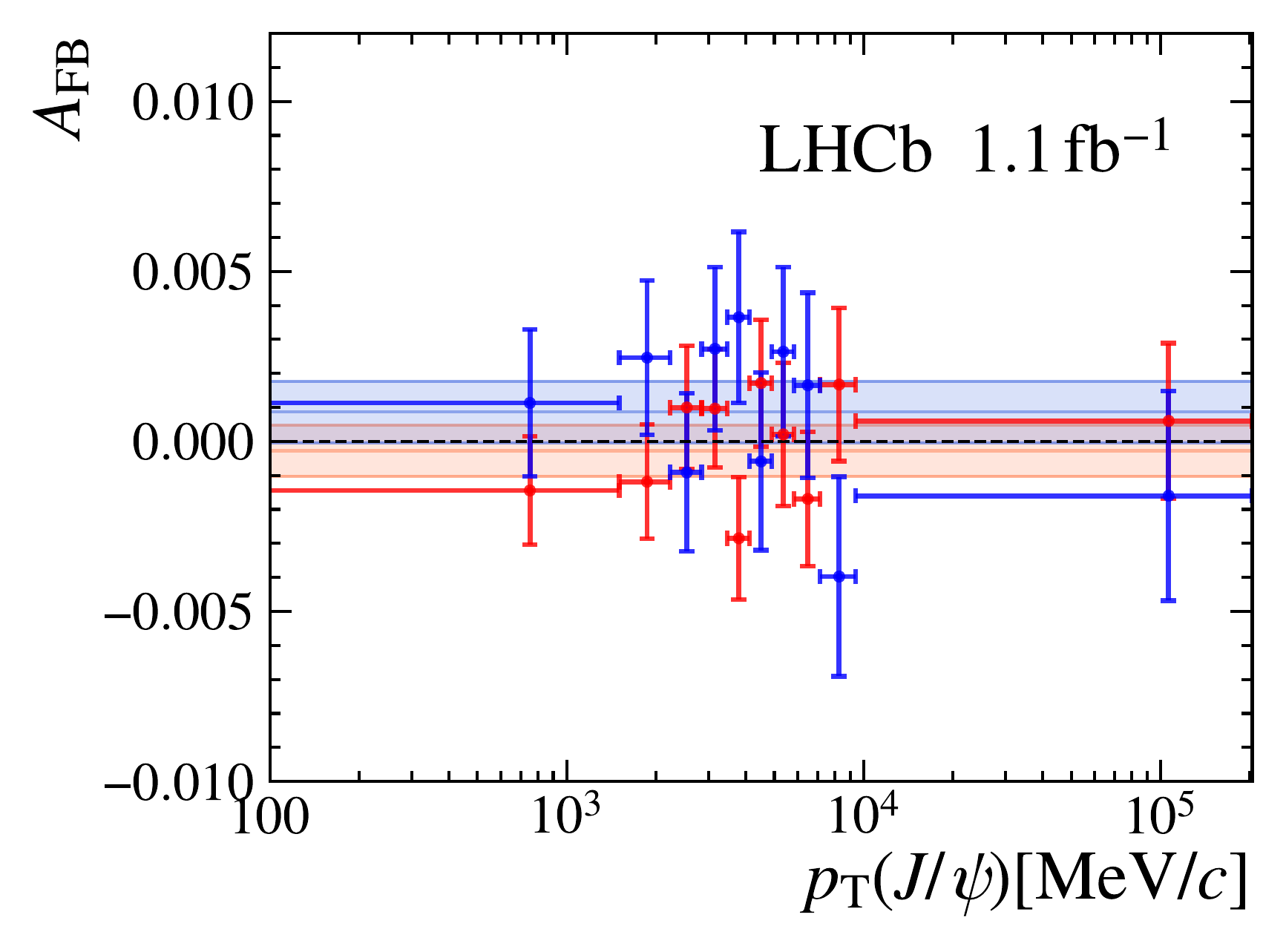}
    \includegraphics[width=0.45\textwidth]{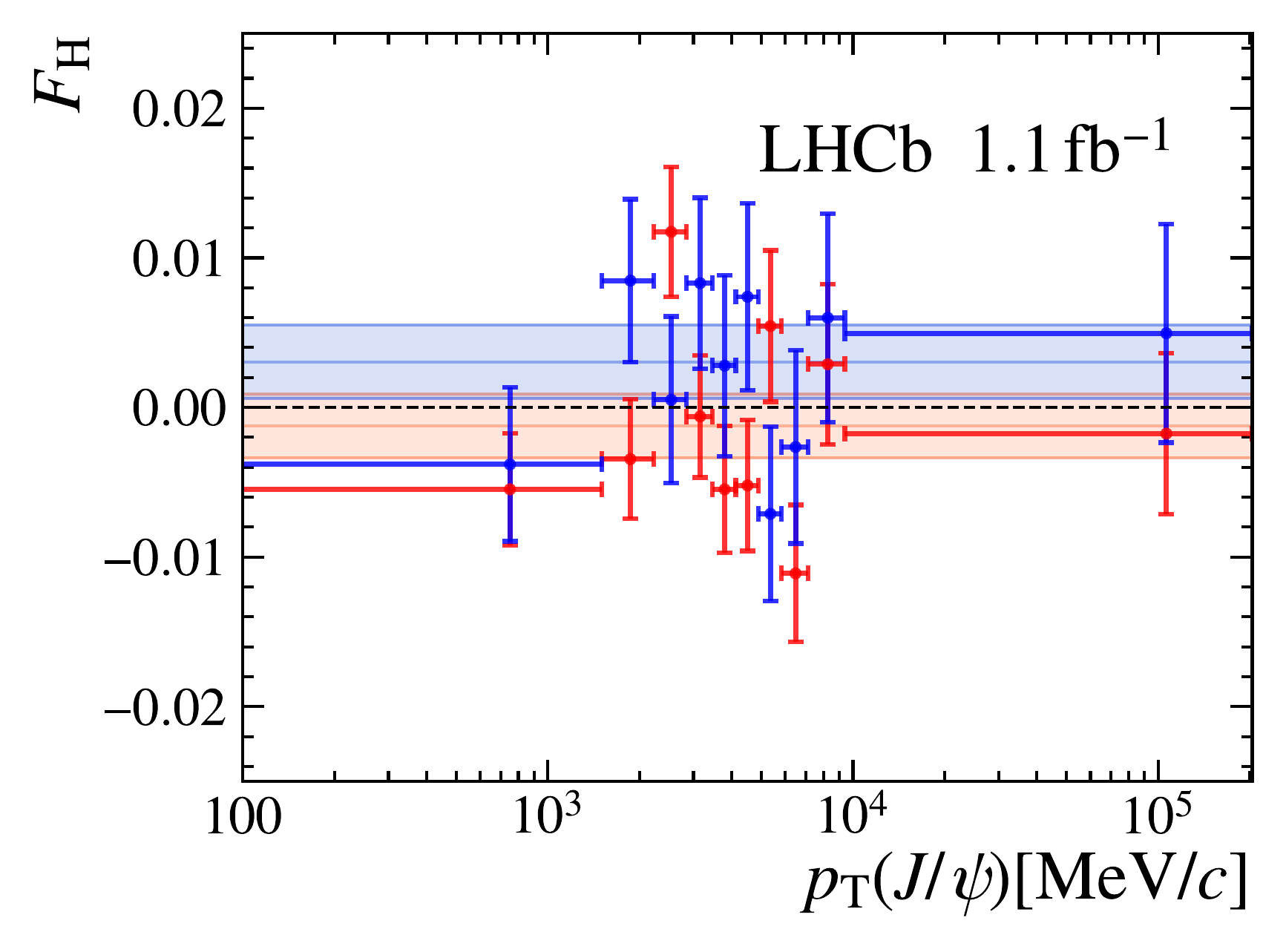}\\ 
    \includegraphics[width=0.45\textwidth]{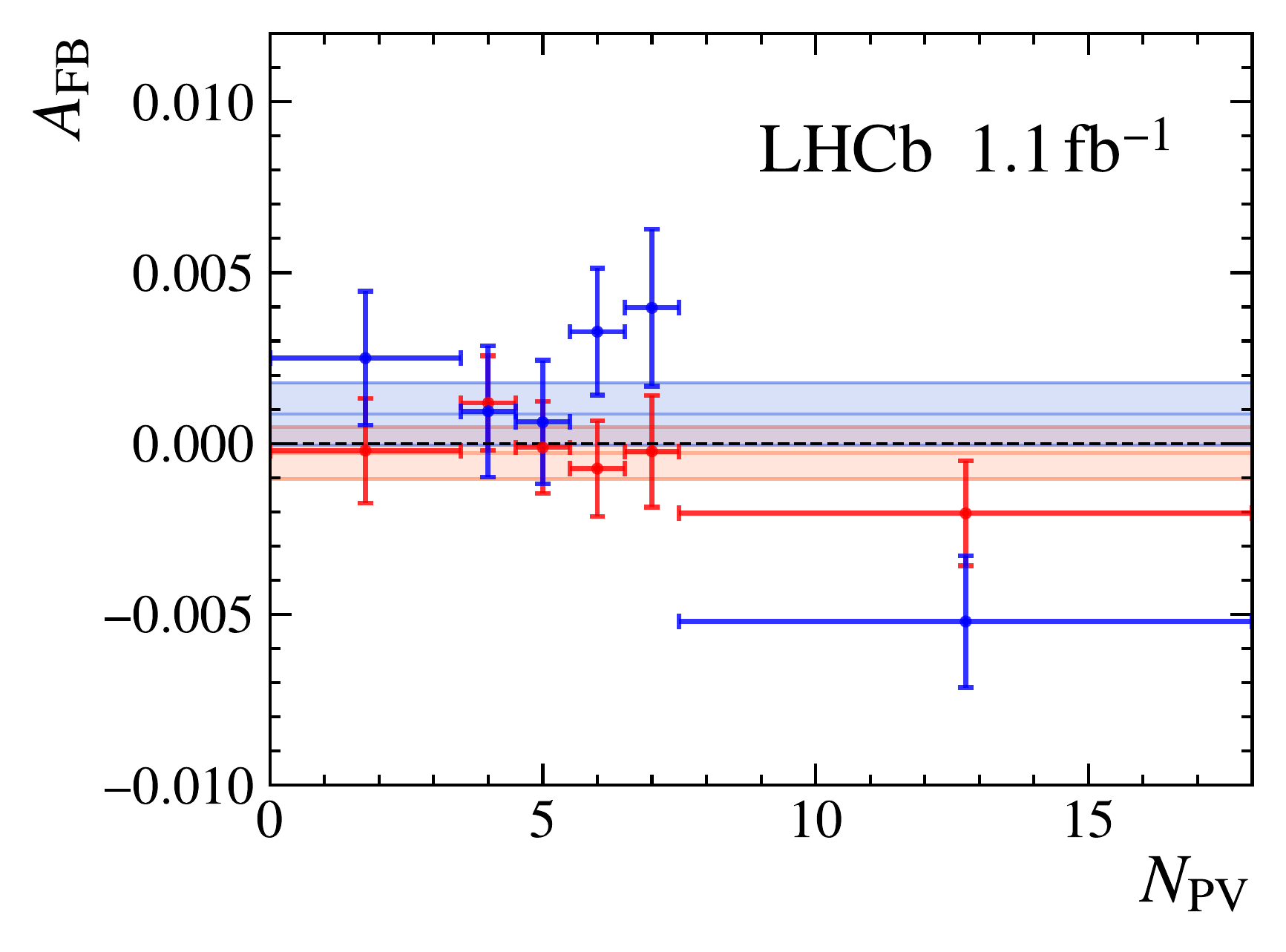}
    \includegraphics[width=0.45\textwidth]{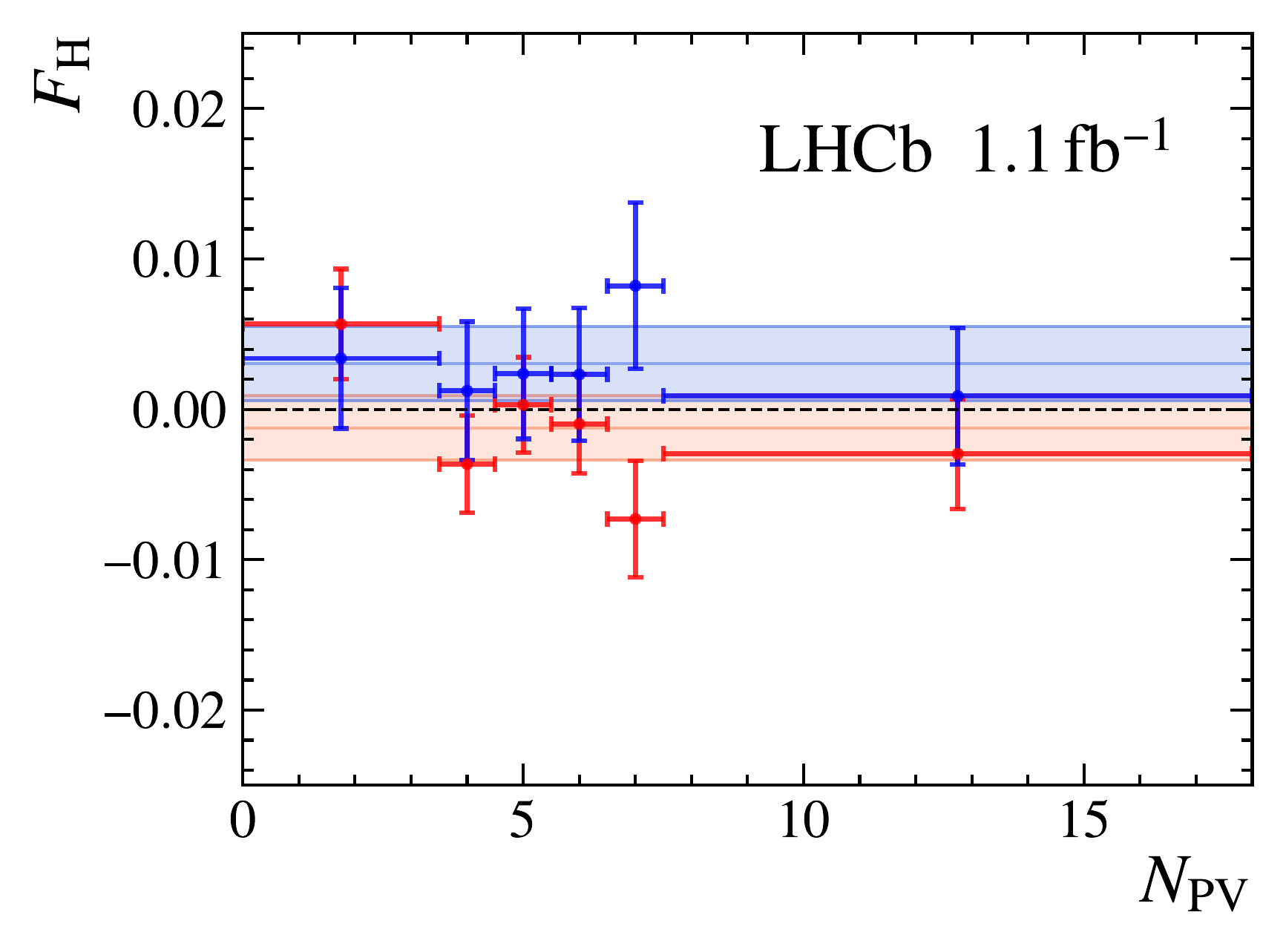}\\    
    
    \caption{Measured values of $A_{\rm FB}$ and $F_{\rm H}$  in ten equipopulated bins of (top)~$\theta(\mu^+\mu^-)$, (middle)~$p_{\rm T}(J/\psi)$, and (bottom)~$N_{\mathrm{PV}}$, for the \emph{MagDown} (red) and \emph{MagUp} (blue) magnet configurations. Only statistical uncertainties are shown. Horizontal bands indicate the corresponding values obtained from the integrated datasets, including statistical and systematic uncertainty contributions.}
     \label{fig:fits_in_bins_block7}
\end{figure}

\begin{figure}[tb]
    \centering
    \includegraphics[width=0.80\textwidth]{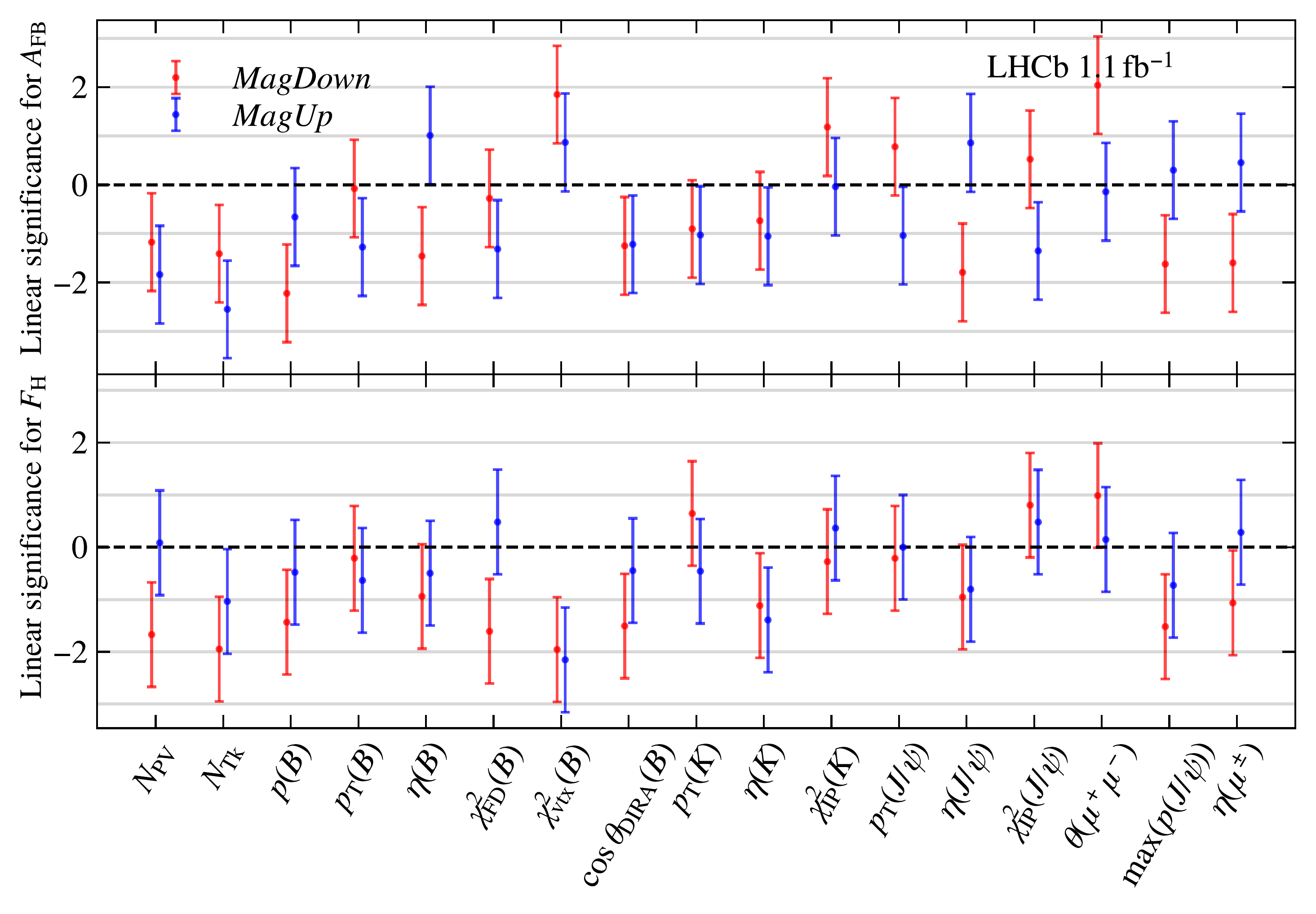}\\[1.5ex]
    \includegraphics[width=0.80\textwidth]{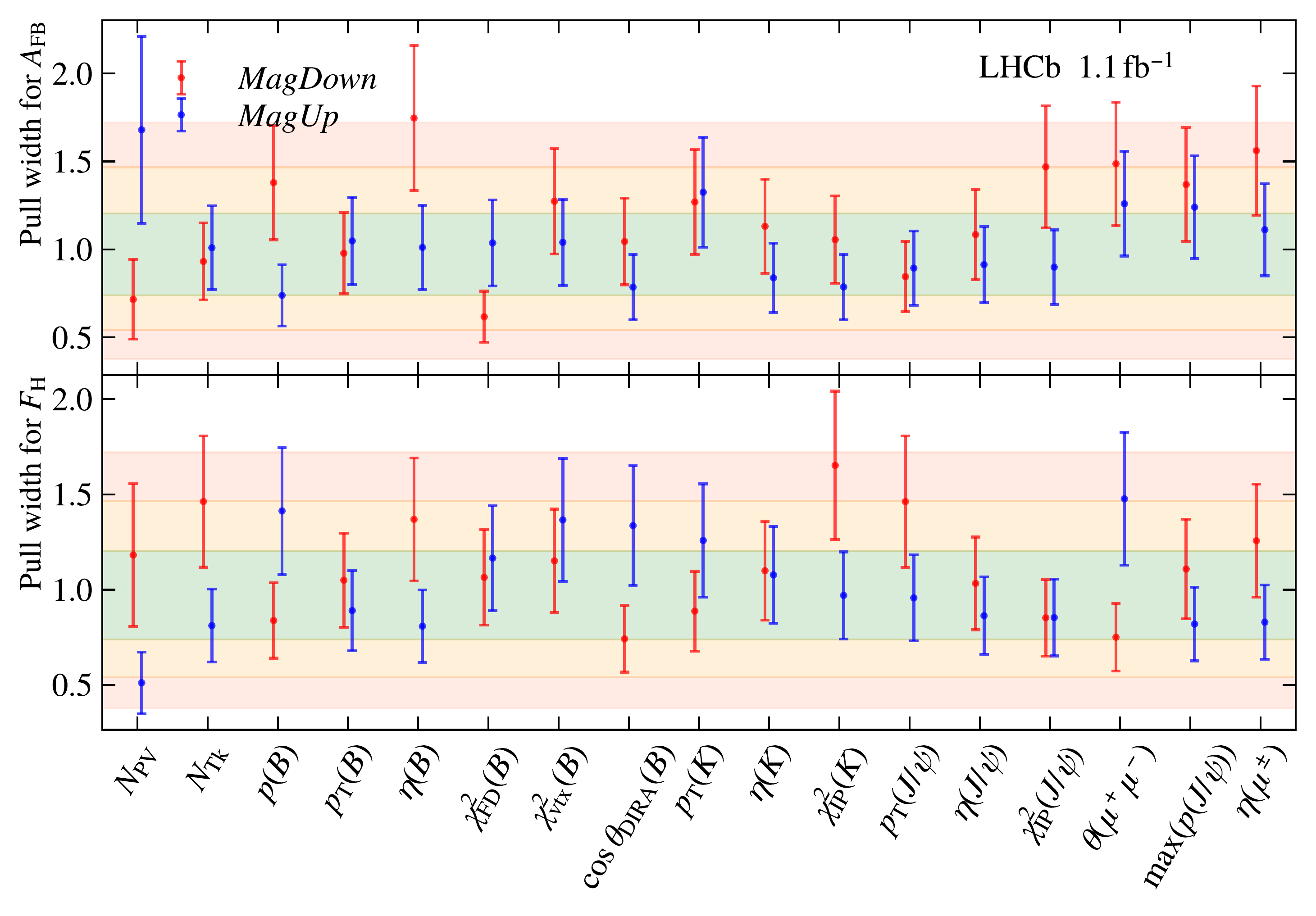}
    \caption{
        (Top) Significance of the linear parameter obtained from a $\chi^2$ fit of the differential variation of the observables $A_{\rm FB}$ and $F_{\rm H}$ as a function of all 17 variables considered.
        (Bottom) Standard deviation of the pull distribution obtained for the observables $A_{\rm FB}$ and $F_{\rm H}$   across the same 17 variables. The green, orange, and red bands indicate the 68.27\%, 95.45\%, and 99.73\% confidence intervals for the width of the pull distribution, derived from pseudoexperiments in which pseudodata is randomly split into ten equally populated bins.
    }
    \label{fig:diff_and_pulls}
\end{figure}

The data in each polarity configuration is also further divided into ten equally populated bins according to the 17 variables previously discussed. The results for some example variables are shown in Fig.~\ref{fig:fits_in_bins_block7}, namely the 
 dimuon opening angle, $\theta(\mu^+\mu^-)$, the \jpsi transverse momentum, and $N_{\mathrm{PV}}$, for both the \MagDown and \MagUp datasets. To assess whether any trend across these variables could affect the final, integrated result,  two metrics are developed. The first is the linear coefficient obtained when fitting the results across a given variable. The significance of the linear coefficient from zero is shown for all variables in  Fig.~\ref{fig:diff_and_pulls} (top). 
 
 As this is only sensitive to linear trends, a second more generic metric is defined via
\begin{equation}\label{eq:pull}
    p(\mathit{O}|_i^m) = \frac{(\mathit{O}|_i^m - \mathit{O}|_{\mathrm{int}}^m)}{\sigma(\mathit{O}|_i^m)} ,
\end{equation}
where $\mathit{O}|_i^m$ and $\sigma(\mathit{O}|_i^m)$ denote the central value and uncertainty for the angular observable $\mathit{O}$ in  bin $i$ of polarity configuration $m$, and $\mathit{O}|_{\mathrm{int}}^m$ is the corresponding central value obtained without binning.
If there is no dependence of the angular coefficients on the variables in question, and the obtained uncertainties are correct, then the pull defined in Eq.~\ref{eq:pull} should follow a normal distribution with a width of unity. 
The requirement on the width is overly conservative, as the uncertainty in the numerator includes only the statistical component, and the leading systematic uncertainty arising from the limited simulation size is also entirely uncorrelated between bins. 

The corresponding results are shown in Fig.~\ref{fig:diff_and_pulls}~(bottom). No evidence of under-coverage is observed. In other words, given the size of the dataset examined, there do not appear to be effects that vary as a function of these variables which are significantly larger than the existing statistical and leading systematic uncertainty.

\begin{figure}[tb]
    \centering
    \includegraphics[width=0.49\textwidth]{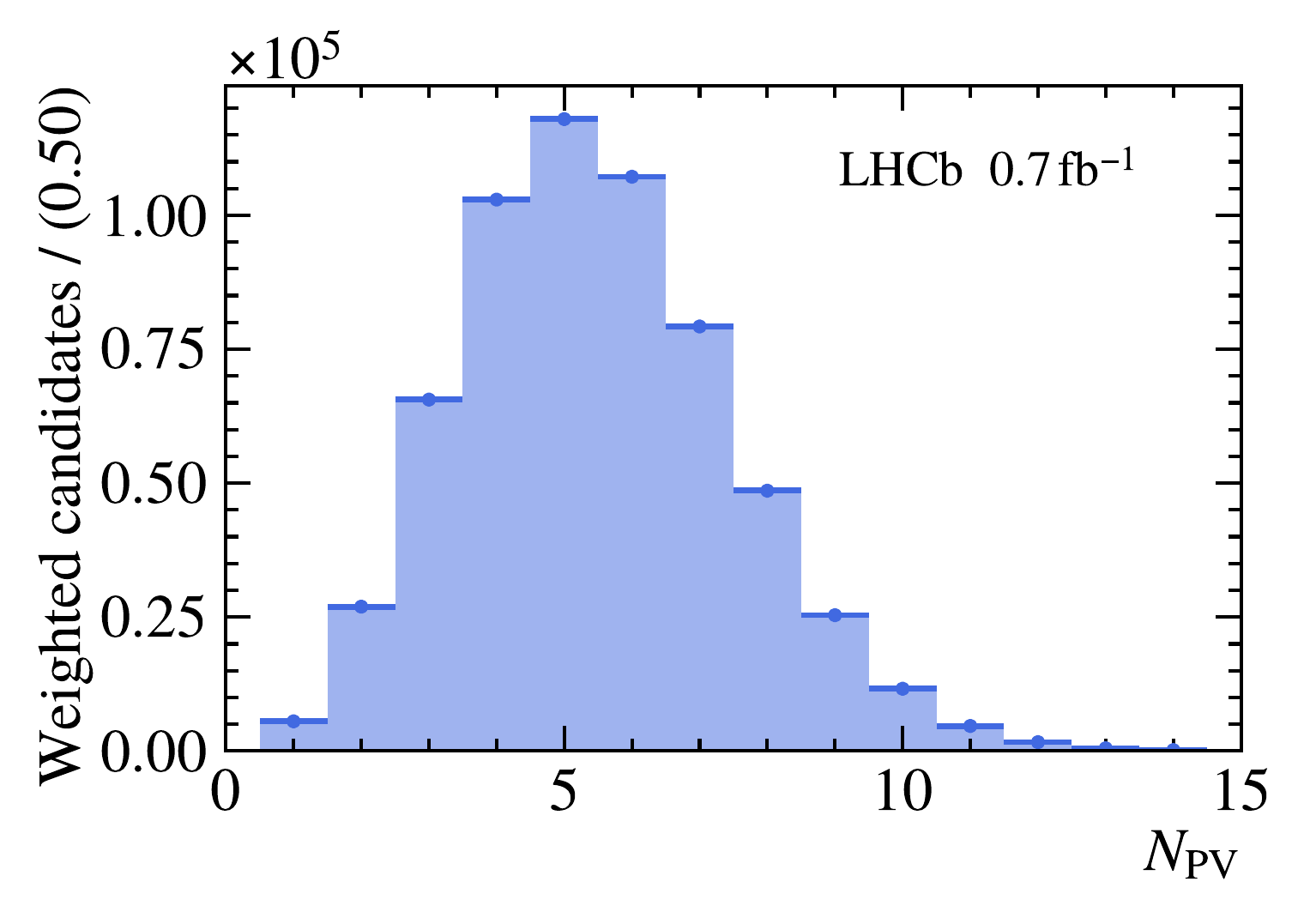}
    \includegraphics[width=0.49\textwidth]{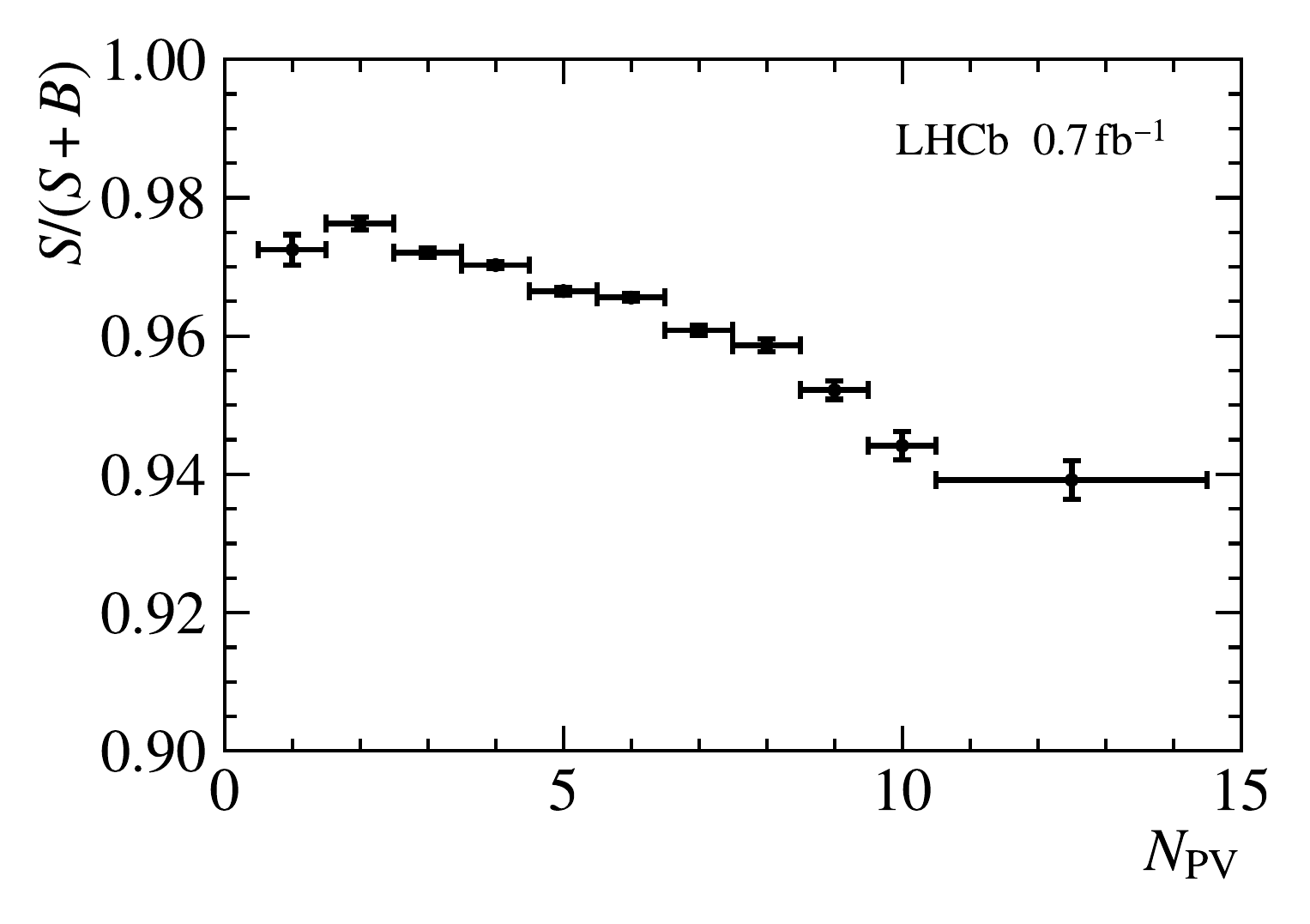}\\    
    \includegraphics[width=0.49\textwidth]{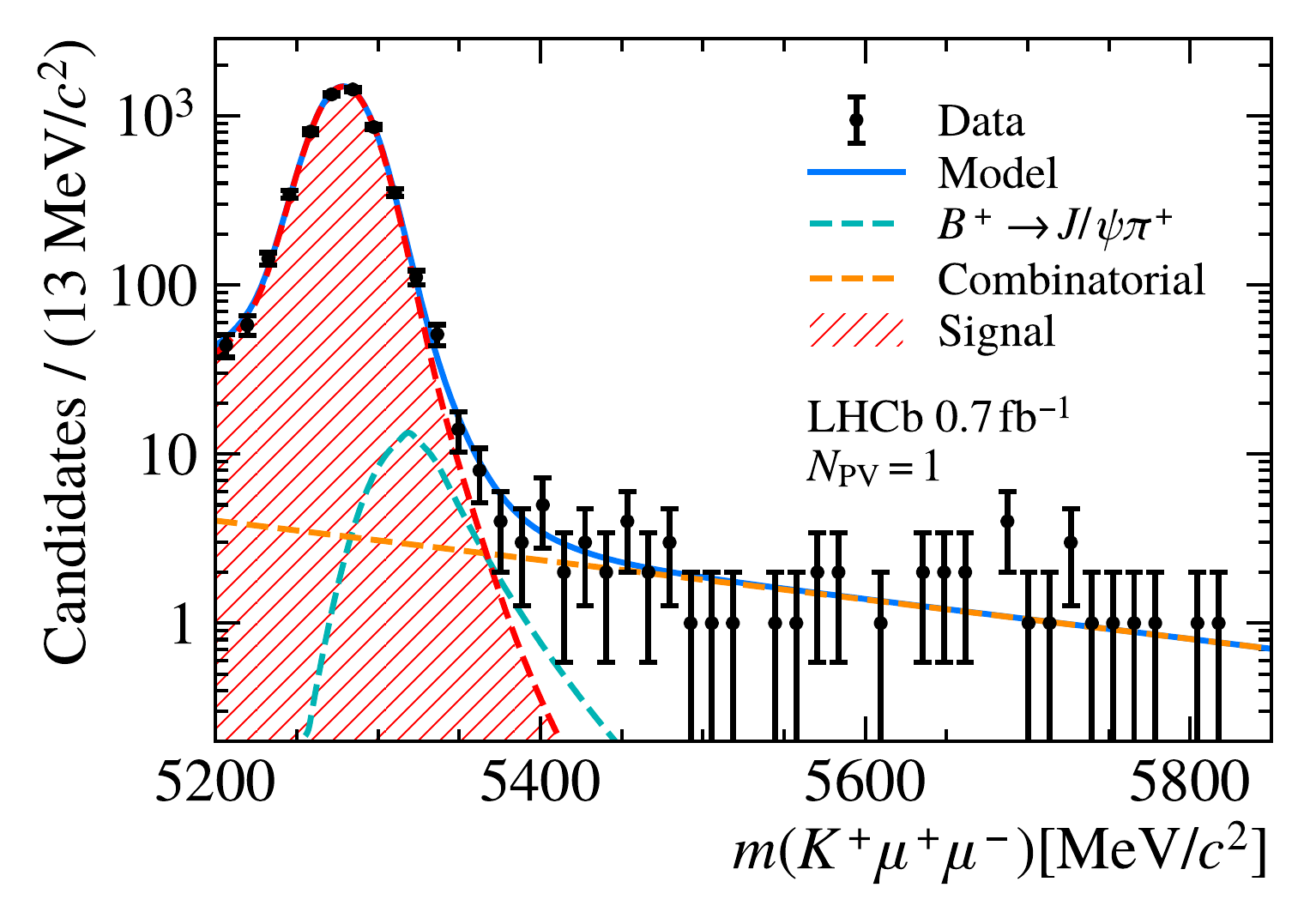}
    \includegraphics[width=0.49\textwidth]{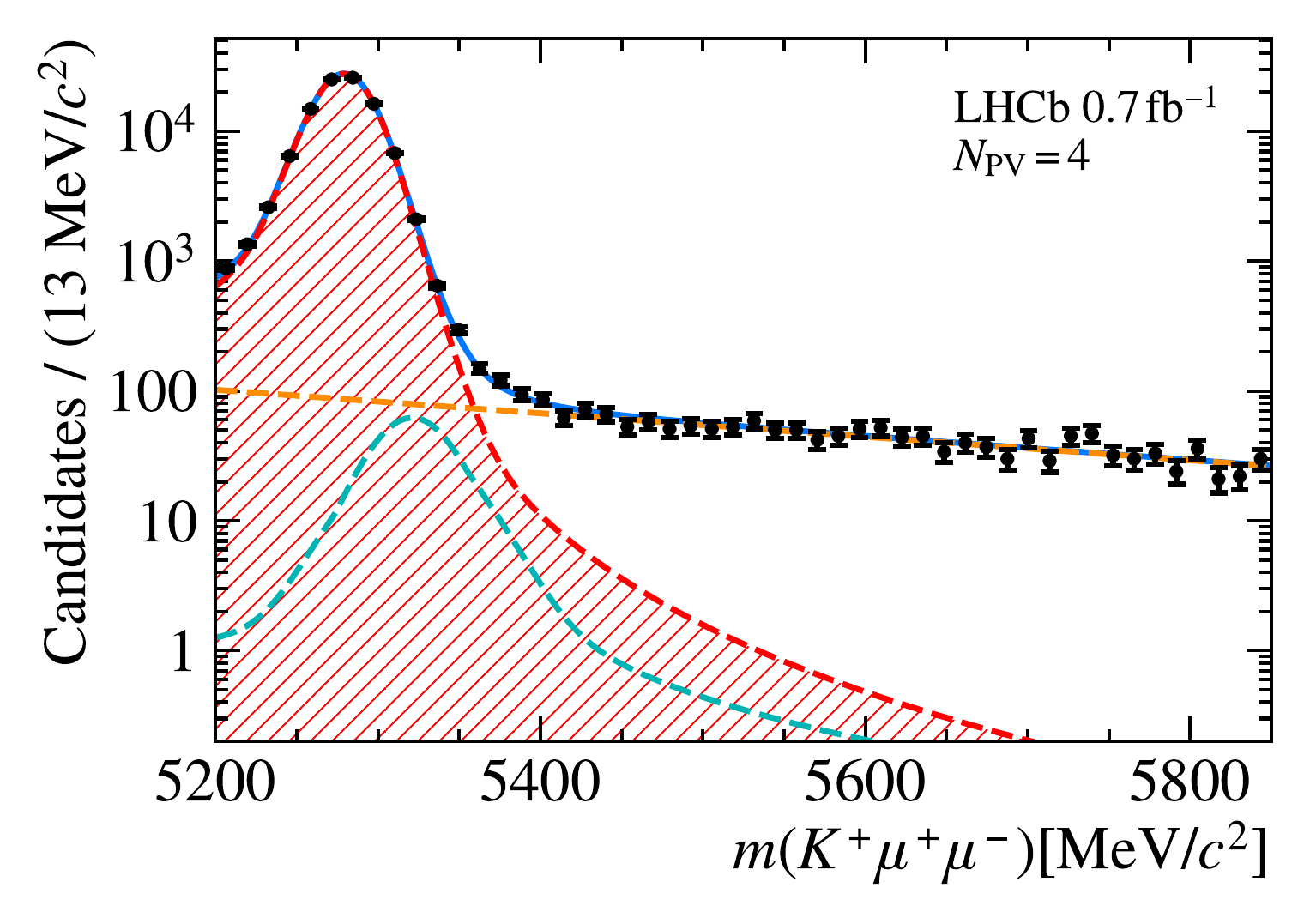}\\
    \includegraphics[width=0.49\textwidth]{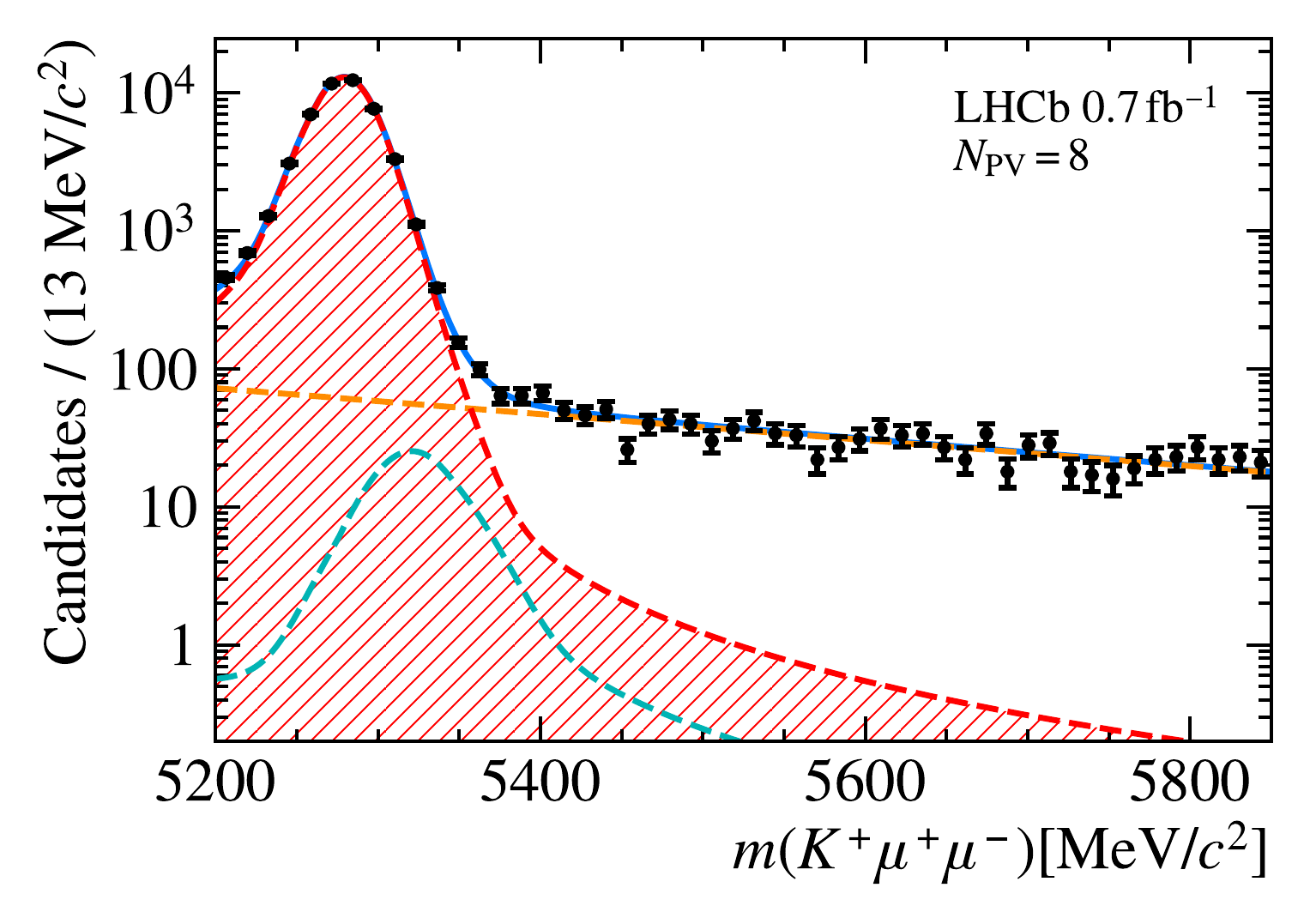}
    \includegraphics[width=0.49\textwidth]{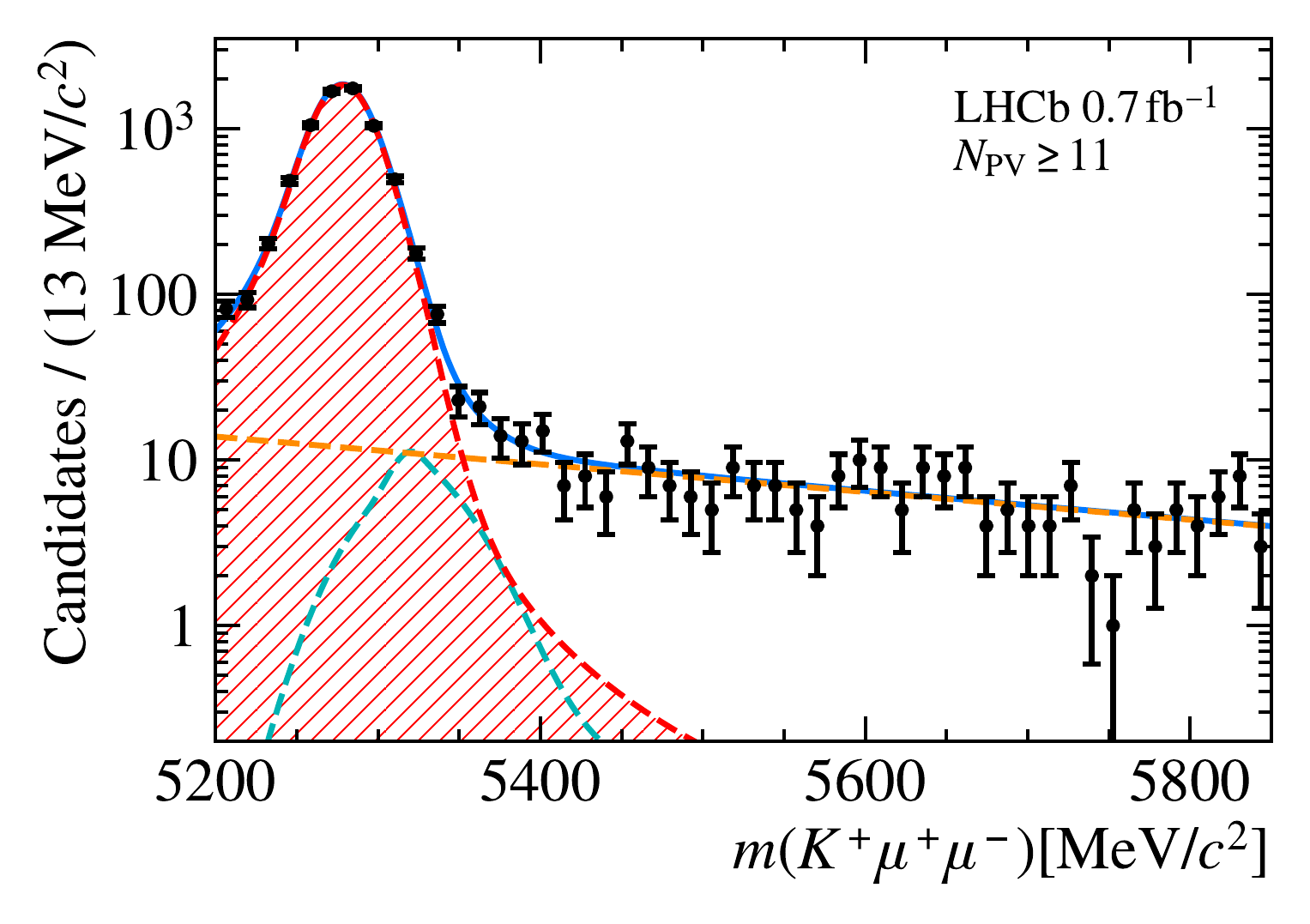}\\
    \caption{(Top left) Background-subtracted distribution of $N_{\mathrm{PV}}$ in \emph{MagDown} data. (Top right)~Signal fraction integrated over the fit mass window as a function of $N_{\mathrm{PV}}$ in \emph{MagDown} data. (Middle and bottom panels) Projections of fits onto 
    $m(K^+\mu^+\mu^-)$ for \emph{MagDown} data in four representative bins of $N_{\mathrm{PV}}$.
}
    \label{fig:nPVs}
\end{figure}

Finally, given that part of the scope of this work is to validate the response of the \lhcb Upgrade~I detector in the new, higher-luminosity environment, Fig.~\ref{fig:nPVs} shows the number of candidates and the signal purity as a function of $N_{\mathrm{PV}}$.
The signal purity is defined as the ratio $S/(S+B)$, where $S$ and $B$ correspond to the numbers of signal and background candidates extracted from the mass fits. Mass projections for $N_{\mathrm{PV}} = 1, 4, 8$, and $N_{\mathrm{PV}} \geq 11$ are also shown in the figure for reference.

\section{Discussion and conclusion}
\label{sec:Conclusion}

In the \qsq-region considered in this work, the dominance of the  \Jpsi resonance means the angular observables  \afb and \fh 
are expected to be consistent with zero, assuming the detector response is correctly described in simulation. 
This expectation is confirmed by the measurements obtained in both the \MagDown and \MagUp magnet configurations, as well as in their average. Statistical and systematic uncertainties contribute in roughly equal proportions to the total uncertainty, with the latter playing a more significant role for \fh than for \afb. This is evident in the breakdown of systematic contributions shown in Table~\ref{tab:syst}: while the dominant systematic uncertainty for \afb arises from the limited size of the simulated samples, \fh is also significantly affected by the choice of alternative schemes for the extraction of \wkin. This difference is not unexpected, as detector- or background-induced effects are less likely to bias \afb, given that such effects would require a charge-asymmetric response to muons.  

Extrapolating the Run~1 measurement of \afb and \fh in \decay{\Bu}{\Kp\mup\mun} decays~\cite{LHCB-PAPER-2014-007} to the end of Run~3, assuming a 100\% increase in the $b$-quark production cross-section due the increased centre-of-mass energy~\cite{LHCb-PAPER-2016-031}, and a total integrated luminosity of 23~\invfb, the expected statistical uncertainties on \afb and \fh in the range $1.1 < \qsq < 6.0$ \gevgevcccc are approximately 0.005 and 0.01, respectively.
These are five to ten times larger than the size of the systematic uncertainties observed on the \KJpsi angular coefficients in this work.  
Furthermore, the systematic uncertainty quoted here is expected to improve in the future, as the leading contributions arise from either limited size of the simulation samples or the current level of understanding of the detector response, both of which should improve as more data are collected. 
The extracted angular coefficients also exhibit good stability when the analysis is restricted to specific subregions of the detector, with, for example, no coherent discrepancies observed between opposite kaon charges across different magnet polarities (Fig.~\ref{fig:kaoncharge}).

Similar conclusions are drawn from the study of the dependence of \afb and \fh on 17 detector-response and kinematic variables, as shown in Fig.~\ref{fig:fits_in_bins_block7} and Fig.~\ref{fig:diff_and_pulls} (top). A linear $\chi^2$ fit to the differential measurements in each variable reveals no significant deviation from zero that is consistent across both magnet polarities. The largest deviations observed reach at most approximately two standard deviations, which is compatible with statistical fluctuations. 
The slight prevalence of negative slopes across these variables should not be interpreted as evidence of a systematic trend, since the linear coefficients extracted from different variable subgroups are correlated. For example, $N_{\mathrm{PV}}$ and $N_{\mathrm{Tk}}$ share a dependence on event multiplicity, while $\cos\theta_{\mathrm{DIRA}}$ and  $\chi^2_{\mathrm{FD}}$ are all related to the displacement of the \Bu candidate.

An alternative interpretation of the results, based on the pull metric defined in Eq.~\ref{eq:pull}, leads to a similar conclusion, as shown in Fig.~\ref{fig:diff_and_pulls} (bottom): the distribution of pull values across the different variables is compatible with expectations from statistical fluctuations. The stability observed across different detector-response and kinematic variables, including those whose underlying distributions are expected to differ significantly between the \KJpsi and \btosmumu modes, such as $p_{T}(\Jpsi)$ and $\theta(\mu^+\mu^-)$, as shown in Fig.~\ref{fig:fits_in_bins_block7} (top and middle), is reassuring. 
The quality of the detector-response modelling relevant for \btosmumu analyses is further evidenced by the consistent results obtained when tightening both the PID and BDT selections to values more representative of those typically used in \btosmumu analyses.

The instantaneous luminosity for the data used in this analysis is approximately five times higher than that of data collected with the previous \lhcb detector.
Figure~\ref{fig:nPVs}~(top left and right) shows the distributions of $N_{\mathrm{PV}}$ and the signal purity as a function of $N_{\mathrm{PV}}$, respectively. Example mass-fit projections for different values of $N_{\mathrm{PV}}$ are shown in the remaining two rows of the same figure.
The signal purity varies by less than 10\% across a wide range of pile-up conditions, ranging from 1 to 14 reconstructed primary vertices per event, with little to no degradation observed in the $B^+$ mass resolution. This, together with the stability of the extracted angular coefficients across different $N_{\mathrm{PV}}$ bins, as shown in Fig.~\ref{fig:fits_in_bins_block7}~(bottom), highlights the robustness of the \lhcb Upgrade~I detector and reconstruction, even under significantly more challenging pile-up conditions.

In summary, this work presents a precise measurement of the angular distribution in \KJpsi decays using 1.1\invfb of data collected in October 2024, with the results showing excellent agreement with theoretical predictions. The analysis is performed on data taken at an instantaneous luminosity of $2 \times 10^{33}~\mathrm{cm}^{-2}~\mathrm{s}^{-1}$, representative of the running conditions in 2025 and those expected in 2026. It marks the first complete physics analysis of a $b$-hadron decay using the upgraded \lhcb detector~(Upgrade I) and highlights the detector’s performance across varying levels of pile-up conditions.
The \KJpsi mode features a final state similar to that of \btosmumu channels, which form a central part of the \lhcb physics programme. Likewise, the \KJpsi lepton helicity angle shares a comparable functional form with the helicity angles in those modes.
The systematic uncertainties of this measurement are significantly smaller than the statistical uncertainties expected for \btosmumu modes. This, combined with the observed stability of the extracted results across a range of detector-response and kinematic variables, as well as under varying selection criteria, provides a robust indication that the response of the \lhcb Upgrade~I detector is understood to a level of precision sufficient for the reliable extraction of angular coefficients in both \btosmumu and \btodmumu transitions.

\section*{Acknowledgements}
%
%
\noindent We express our gratitude to our colleagues in the CERN
accelerator departments for the excellent performance of the LHC. We
thank the technical and administrative staff at the LHCb
institutes.
We acknowledge support from CERN and from the national agencies:
ARC (Australia);
CAPES, CNPq, FAPERJ and FINEP (Brazil); 
MOST and NSFC (China); 
CNRS/IN2P3 and CEA (France);
BMFTR, DFG and MPG (Germany);
INFN (Italy); 
NWO (Netherlands); 
MNiSW and NCN (Poland); 
MEC/IFA (Romania);
MICIU and AEI (Spain);
SNSF and SER (Switzerland); 
NASU (Ukraine); 
STFC (United Kingdom); 
DOE NP and NSF (USA).
We acknowledge the computing resources that are provided by ARDC (Australia), 
CBPF (Brazil),
CERN, 
IHEP and LZU (China),
IN2P3 (France), 
KIT and DESY (Germany), 
INFN (Italy), 
SURF (Netherlands),
Polish WLCG (Poland),
IFIN-HH (Romania), 
PIC (Spain), CSCS (Switzerland), 
GridPP (United Kingdom),
and NSF (USA).
We are indebted to the communities behind the multiple open-source
software packages on which we depend.
Individual groups or members have received support from
RTP (Australia),
Key Research Program of Frontier Sciences of CAS, CAS PIFI, CAS CCEPP (China);
Minciencias (Colombia);
EPLANET, Marie Sk\l{}odowska-Curie Actions, ERC and NextGenerationEU (European Union);
A*MIDEX, ANR, IPhU and Labex P2IO, and R\'{e}gion Auvergne-Rh\^{o}ne-Alpes (France);
Alexander-von-Humboldt Foundation (Germany);
ICSC (Italy); 
Severo Ochoa and Mar\'ia de Maeztu Units of Excellence, GVA, XuntaGal, GENCAT, InTalent-Inditex and Prog.~Atracci\'on Talento CM (Spain);
the Leverhulme Trust, the Royal Society and UKRI (United Kingdom).




\clearpage
\addcontentsline{toc}{section}{References}
\bibliographystyle{LHCb}
\bibliography{main,standard,LHCb-PAPER,LHCb-CONF,LHCb-DP,LHCb-TDR}

\newpage
\centerline
{\large\bf LHCb collaboration}
\begin
{flushleft}
\small
R.~Aaij$^{38}$\lhcborcid{0000-0003-0533-1952},
A.S.W.~Abdelmotteleb$^{58}$\lhcborcid{0000-0001-7905-0542},
C.~Abellan~Beteta$^{52}$\lhcborcid{0009-0009-0869-6798},
F.~Abudin\'en$^{58}$\lhcborcid{0000-0002-6737-3528},
T.~Ackernley$^{62}$\lhcborcid{0000-0002-5951-3498},
A. A. ~Adefisoye$^{70}$\lhcborcid{0000-0003-2448-1550},
B.~Adeva$^{48}$\lhcborcid{0000-0001-9756-3712},
M.~Adinolfi$^{56}$\lhcborcid{0000-0002-1326-1264},
P.~Adlarson$^{86}$\lhcborcid{0000-0001-6280-3851},
C.~Agapopoulou$^{14}$\lhcborcid{0000-0002-2368-0147}
% ...

{\footnotesize \it

% ...
$^{14}$Universit{\'e} Paris-Saclay, CNRS/IN2P3, IJCLab, Orsay, France\\
% ...
$^{38}$Nikhef National Institute for Subatomic Physics, Amsterdam, Netherlands\\
% ...
$^{48}$Instituto Galego de F{\'\i}sica de Altas Enerx{\'\i}as (IGFAE), Universidade de Santiago de Compostela, Santiago de Compostela, Spain\\
% ...
$^{52}$Physik-Institut, Universit{\"a}t Z{\"u}rich, Z{\"u}rich, Switzerland\\
% ...
$^{56}$H.H. Wills Physics Laboratory, University of Bristol, Bristol, United Kingdom\\
% ...
$^{58}$Department of Physics, University of Warwick, Coventry, United Kingdom\\
% ...
$^{61}$School of Physics and Astronomy, University of Glasgow, Glasgow, United Kingdom\\
$^{62}$Oliver Lodge Laboratory, University of Liverpool, Liverpool, United Kingdom\\
% ...
$^{70}$Syracuse University, Syracuse, NY, United States\\
% ...
$^{86}$Department of Physics and Astronomy, Uppsala University, Uppsala, Sweden, associated to $^{61}$\\
% ...
}
\end{flushleft}


\end{document}